\documentclass{aastex}
\usepackage{emulateapj5}

\newcommand{\n}{\nodata}

\def\gtrsim{\mathrel{\hbox{\rlap{\hbox{\lower4pt\hbox{$\sim$}}}\hbox{$>$}}}}
\def\lesssim{\mathrel{\hbox{\rlap{\hbox{\lower4pt\hbox{$\sim$}}}\hbox{$<$}}}}

\slugcomment{Accepted for publication in the {\it Astrophysical Journal}.}
\shortauthors{Lister}
\shorttitle{Parsec-Scale Jet Polarization Properties of Active Galactic Nuclei at 43 GHz}

\begin{document}

\title{Parsec-Scale Jet Polarization Properties of a Complete Sample \\
       of Active Galactic Nuclei at 43 GHz}
\author{Matthew  L. Lister\altaffilmark{1}}
\affil{National Radio Astronomy Observatory, 520 Edgemont Road, Charlottesville, VA 22903-2454}

\altaffiltext{1}{Former address: Jet Propulsion Laboratory, California
Institute of Technology, MS 238-332, 4800 Oak Grove Drive, Pasadena, CA 91109}

\email{mlister@nrao.edu}

\begin{abstract}
We present results from the highest resolution polarization imaging
survey of a complete sample of extragalactic radio sources carried out
to date. Our sample consists of a statistically complete flat-spectrum
subset of 32 active galactic nuclei (AGN) from the Pearson-Readhead
survey, which has been studied at a variety of wavelengths and
resolutions, including space-VLBI.  Our observations were made with
the VLBA at 43 GHz, where the relatively higher resolution and weaker
opacity effects have allowed us to probe magnetic field structures in
the jets of these AGNs much closer to the central engine than in
previous studies. At 43 GHz, the bulk of the total intensity and
polarized emission in most flat-spectrum AGNs originates from an
unresolved core component located at the extreme end of a faint jet.
The luminosity of the core is positively correlated with the total
source luminosity in soft x-rays, in the optical, and at 5 GHz. The
most strongly polarized cores display electric vectors that are
preferentially aligned with the jet axis, which is consistent with a
strong transverse shock that enhances the perpendicular component of
the jet magnetic field. Sources with highly polarized cores also tend
to have high optical polarizations and flatter overall radio spectra.
Approximately half of the AGNs in our sample display apparently bent
jet morphologies that are suggestive of streaming motions along a
helical path. The straightest jets in the sample tend to display
slower superluminal speeds than those that are significantly bent. Our
observations also show that intrinsic differences in the jet magnetic
field properties of BL Lacertae objects and quasars previously seen on
scales of tens of milliarcseconds are also present in regions much
closer to the base of the jet.

\end{abstract}


\section{Introduction}
Studies of active galactic nuclei (AGNs) with polarization-sensitive
VLBI currently provide us with one of the best means of investigating
the magnetic field structures of outflows from the vicinity of
supermassive black holes. Theoretical and numerical simulations of
these outflows suggest that they are highly collimated and accelerated
to relativistic speeds by magnetic fields generated in a rapidly
rotating accretion disk (see recent review by \citealt{MKU01}). Many
of these outflows are believed to undergo strong shocks that
significantly alter their observed polarization structure (e.g.,
\citealt{HAA85}).  

Past VLBI polarization observations of large AGN samples (e.g.,
\citealt*{CWRG93,GPC00}) have led to important discoveries regarding
relativistic shocks and the underlying magnetic fields of AGN
jets. These surveys have generally been carried out at cm-wavelengths
however, where Faraday effects can be important.  At shorter observing
wavelengths, it is possible to probe regions much closer to the base
of the jet due to the $\lambda^{-1}$ improvement in angular resolution
and lower source opacities.  The higher resolution afforded at
mm-wavelengths is especially important in alleviating the effects of
blending within the synthesized beam, which can dramatically reduce
the observed polarized emission from closely spaced,
orthogonally-polarized regions. Faraday rotation and de-polarization
effects are also greatly diminished in the mm-regime, enabling a more
accurate determination of true magnetic field directions.

We present here the results of the first VLBI polarization study of a
complete sample of flat-spectrum AGNs at a wavelength of $\lambda = 7$ mm ($\nu =
43$ GHz).  The synthesized beam in our images has a typical FWHM
angular size of $\sim 0.2$ milliarcseconds, making this the highest
resolution imaging survey of a complete AGN sample made to
date.  Our sample is drawn from the well-known Pearson-Readhead survey
\citep{PR88}, which has been the subject of numerous studies at a
variety of wavelengths and resolutions, including space-VLBI at 5 GHz
\citep{L-PRI}. The latter data have allowed us to trace weak
steep-spectrum jet emission out to much larger distances from the core
than is possible at 43 GHz alone. 

The paper is laid out as follows. In \S~2 we describe our sample
selection criteria, followed by a discussion of our observations and
data analysis methods in \S~3. In \S~4 we discuss various correlations
between observed properties of our sample, and their implications for
current jet and shock models. These include correlations involving
VLBI core component luminosity and polarization, and the morphology
and polarization properties of components located further down the
jet. We also discuss a correlation between the apparent bending
morphologies and speeds in AGN jets, and demonstrate how many of the
bent jet morphologies in our sample are consistent with low-pitch helical
jet trajectories. In \S~5 we discuss apparent differences in the
intrinsic magnetic field structures of BL Lacertae objects and quasars
at 43 GHz that support previous findings at lower observing
frequencies. We summarize our results in
\S~6.

 Throughout this paper we use a Freidmann cosmology with $H_o =
 100\;h \rm
\; km \; s^{-1} \;Mpc^{-1}$, $h = 0.65$, $q_o = 0.1$, and zero cosmological
constant.  We give all position angles in degrees east of north, and
define the spectral index according to $S_\nu \propto
\nu^\alpha$. All fractional polarizations are quoted in per cent, and
refer to linear polarization only.


\section{Sample selection} 
Our sample is derived from the Pearson-Readhead (PR) survey
\citep{PR88}, which consists of all northern ($\delta > +35\arcdeg$)
AGNs with total 5 GHz flux density greater than 1.3 Jy and galactic
latitude $|b| > 10\arcdeg$. Since the steep-spectrum lobe emission can
still make a significant contribution to the total flux of an AGN at 5
GHz, many of the arcsecond-scale core components in the PR sample are
extremely radio weak ($\lesssim 1 \rm \; mJy$ at 43 GHz). In order to
select a statistically complete core-selected sample that is suitable
for snapshot imaging with the NRAO's\footnote{The National Radio
Astronomy Observatory is a facility of the National Science Foundation
operated under cooperative agreement by Associated Universities, Inc.}
Very Long Baseline Array (VLBA), we used a spectral flatness
criterion $\bar\alpha_{5-15} > -0.4$, where $\bar\alpha_{5-15}$ is a
time-averaged single dish spectral index between 5 and 15 GHz measured
by \cite{AAH92}. This final flat-spectrum Pearson-Readhead sample
(hereafter referred to as the FS-PR sample) was originally defined by
us in a related paper \citep{LTP01}, and contains 21 quasars, 9 BL Lac
objects and 2 radio galaxies (see Table~\ref{genprops}).

\section{Observations and data analysis}
We obtained single-epoch 43 GHz VLBA polarization data on the
entire FS-PR sample with the exception of 0954+556. This unusual
flat-spectrum object has a weak core at high frequencies and an
apparent compact symmetric morphology (e.g., \citealt{WPR94}) at 8.4
GHz (A. P. Marscher, private communication). We also observed four PR
sources that did not make the flat-spectrum cut-off: 0153+744,
0538+498, 0711+356, and 1828+487. We include their images and data
here, but do not consider them in the statistical analyses that
follow. We carried out observations of 28 sources over three separate
VLBA sessions, with approximately 60 minutes integration time devoted
to each source. During the first observing session (1999~Apr~6) only
right-hand circular polarization data were obtained at Owens Valley
due to a receiver failure. No data were obtained with the Brewster
antenna during the 2000~Sep~10 session, also due to a receiver
failure. Data from all ten VLBA antennas were obtained during the
2000~Jan~3 session.

The data were recorded in eight baseband channels (IFs) of 8 MHz
bandwidth using 1-bit sampling.  Both right and left hand
polarizations were recorded simultaneously in IF pairs, giving a total
observing bandwidth of 32 MHz. Snapshot data on seven remaining FS-PR
sources were made available to us from other VLBA programs that
used identical observing parameters (see Table~\ref{mapstats}).

The data were correlated using the VLBA correlator in Socorro, NM, and
subsequent data editing and calibration were performed at the Jet
Propulsion Laboratory using the Astronomical Image Processing System
(AIPS) software supplied by NRAO. Our calibration method followed that
of \cite{LS00}. For each observing session we determined the antenna
polarization leakage factors by running the AIPS task LPCAL on four
program sources and averaging the resulting solutions. The antenna
leakage factors ranged from approximately 1 to 7\%, and had a typical
scatter of $0.7\%$. Based on 37 GHz
flux density measurements of 1803+794 from the Mets\"ahovi observatory
(H. Ter\"asranta, private communication), we estimate our absolute
flux density scaling to be accurate to within $\sim 20\%$.

We determined the absolute electric vector position angle (EVPA)
corrections using measurements of component C4 in 3C~279 at each
observing epoch (see \citealt{T00}). As an additional check on our
calibration, we found good agreement between our integrated EVPA for
our 2000~Sep~10 epoch of 3C~279 and the measured value from the NRAO's
Very Large Array
(VLA)\footnote{http://www.aoc.nrao.edu/$\sim$smyers/calibration/} on
the following day. Based on these comparisons, we estimate that our
measured EVPAs are accurate to within $\sim 5\arcdeg$. 

\begin{figure*}
\centering
\includegraphics[angle=-90, width=5in]{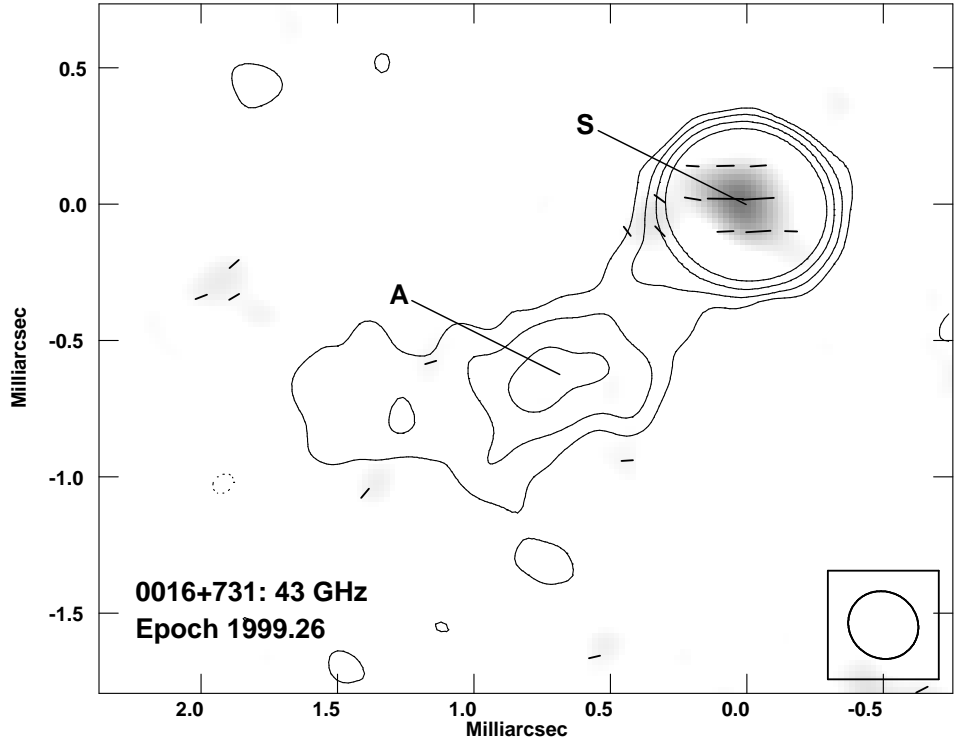}
\caption{\label{map1} VLBA total intensity image of 0016+731
at 43 GHz, with electric vectors superimposed. The gray scale
indicates linearly polarized intensity. }
\end{figure*}

\begin{figure*}
\centering
\includegraphics[angle=-90, width=5in]{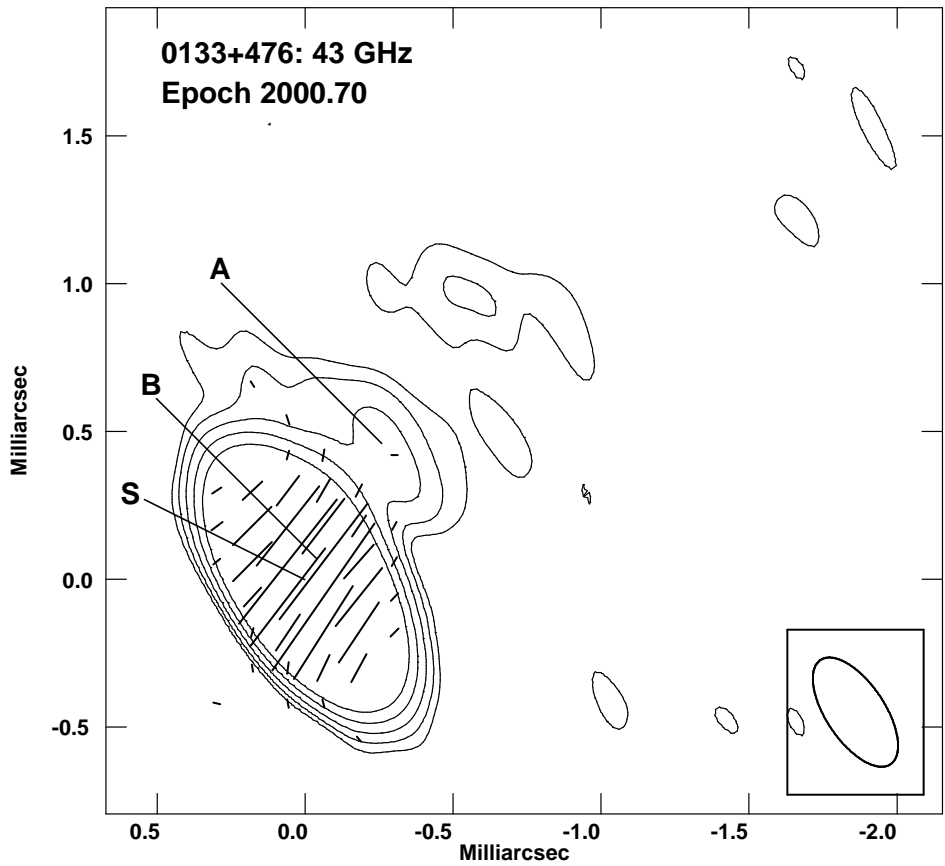}
\caption{\label{map2} VLBA total intensity image of 0133+476 (OC 457)
at 43 GHz, with electric vectors superimposed.  }
\end{figure*}

\begin{figure*}
\centering
\includegraphics[angle=-90, width=5in]{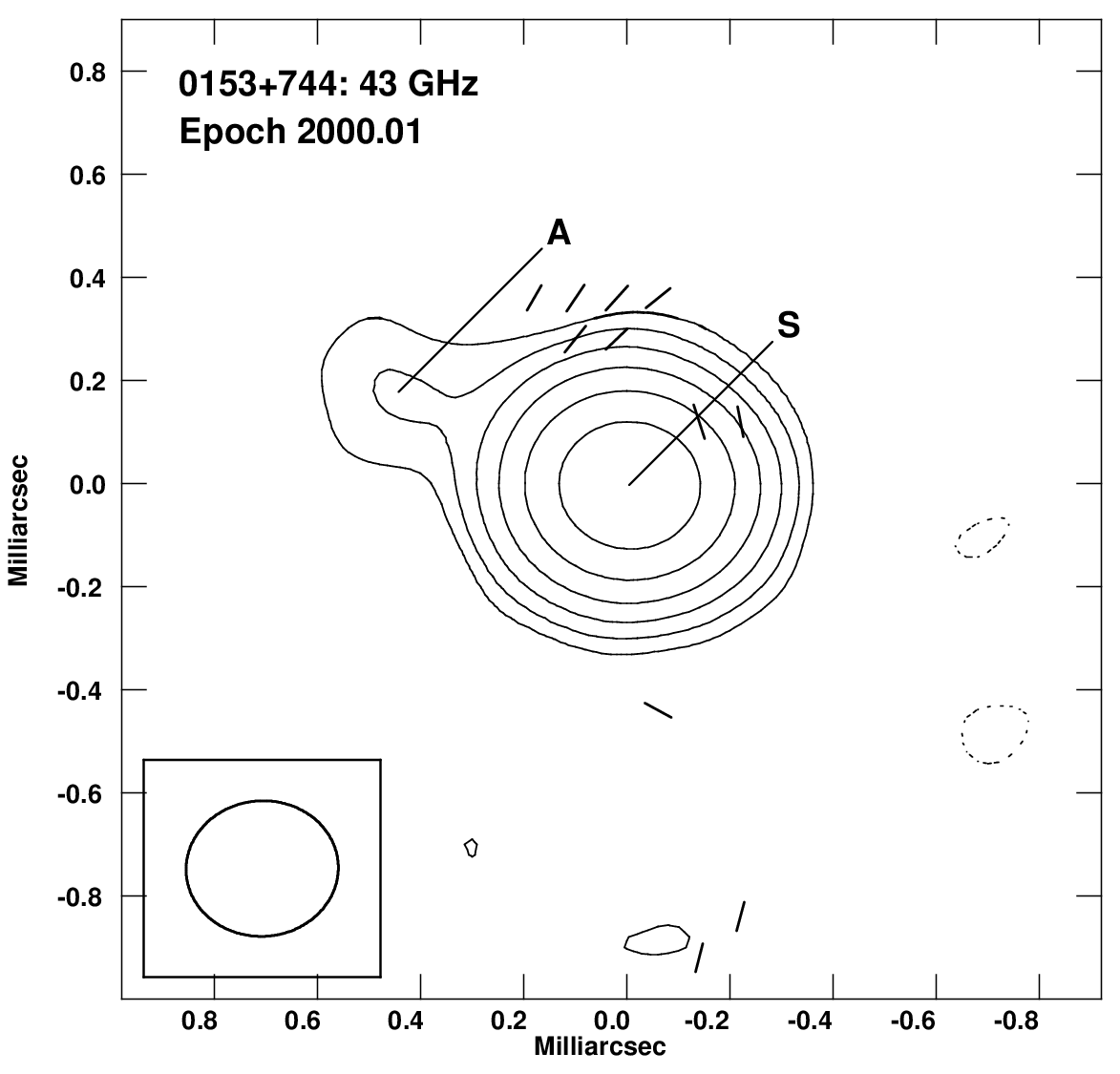}
\caption{\label{map3} VLBA total intensity image of 0153+744
at 43 GHz, with electric vectors superimposed.  }
\end{figure*}

\begin{figure*}
\centering
\includegraphics[angle=-90, width=5in]{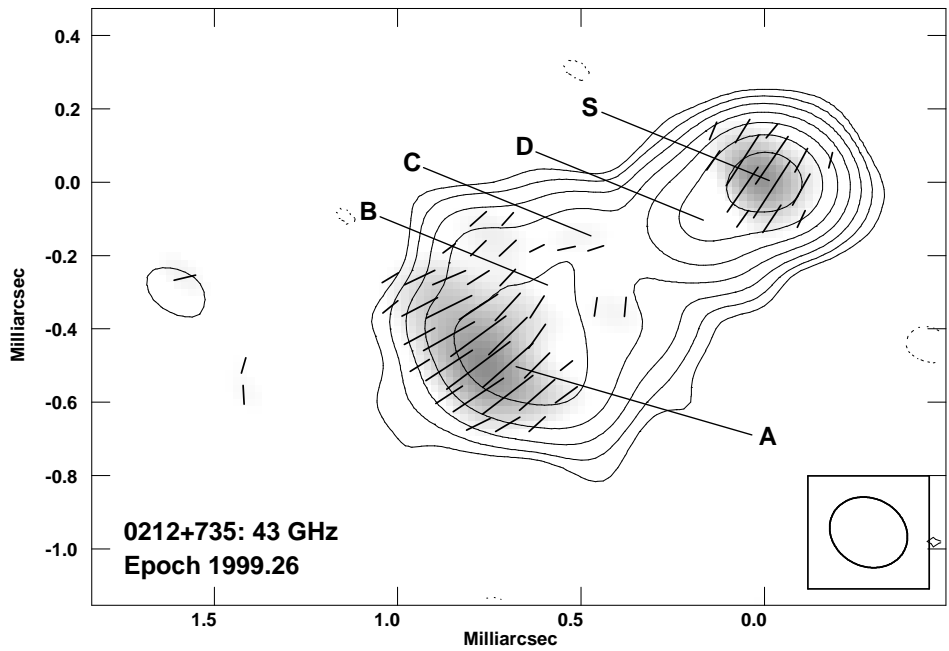}
\caption{\label{map4} VLBA total intensity image of 0212+735 at 43 GHz, with 
electric vectors superimposed. The gray scale
indicates linearly polarized intensity.  }
\end{figure*}

\begin{figure*}
\centering
\includegraphics[width=4in]{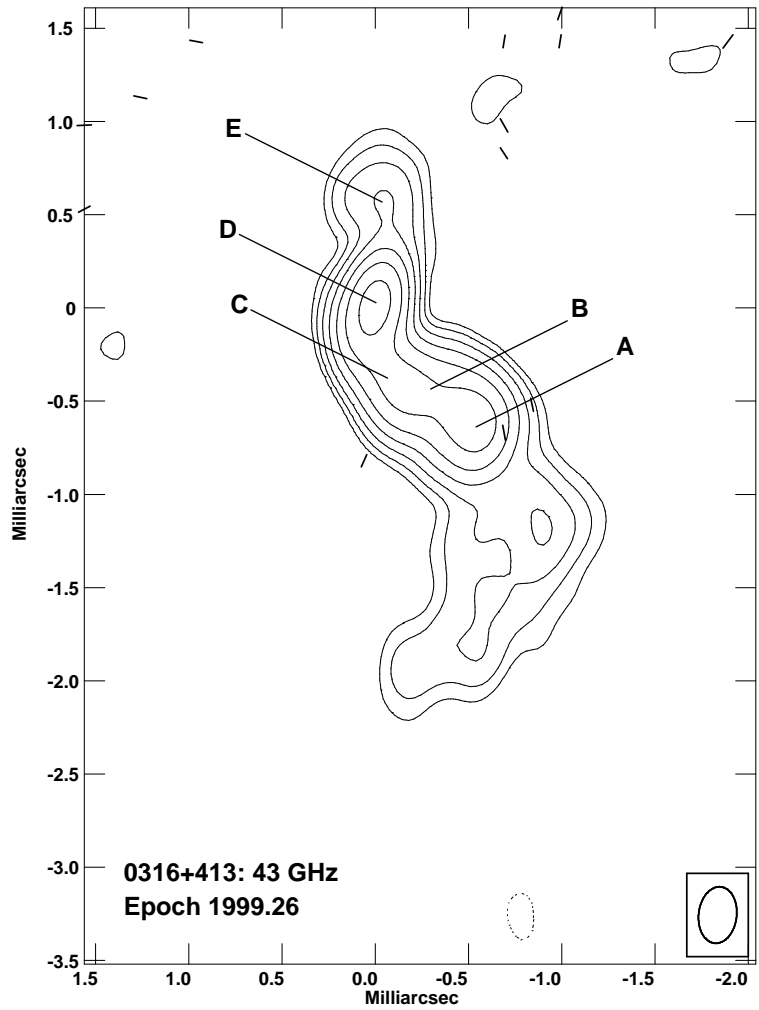}
\caption{\label{map5} VLBA total intensity image of 0316+413 (3C~84) at 43 GHz,
 with electric vectors superimposed.  }
\end{figure*}

\begin{figure*}
\centering
\includegraphics[width=3.5in]{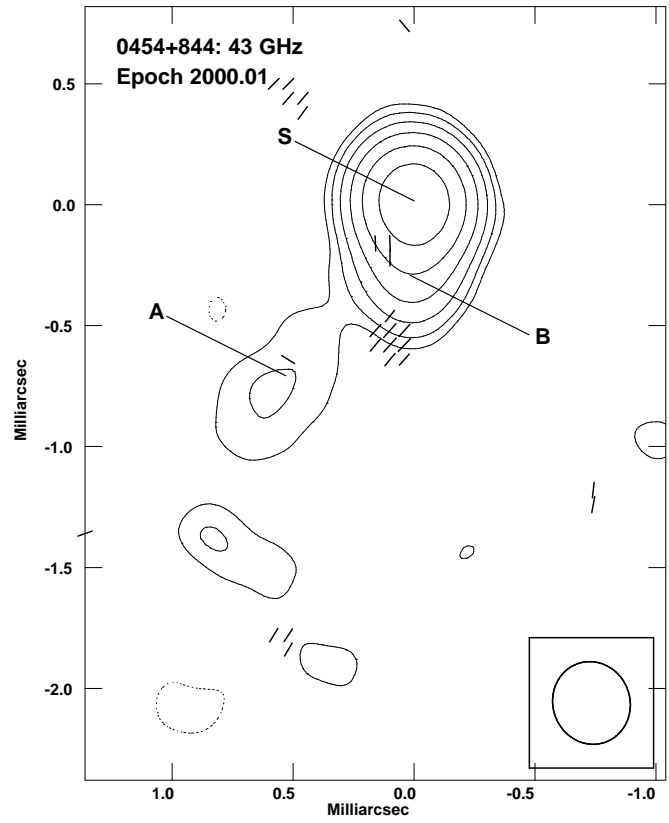}
\caption{\label{map6} VLBA total intensity image of 0454+844 at 43 GHz, with 
electric vectors superimposed.   }
\end{figure*}

\begin{figure*}
\centering
\includegraphics[angle = -90, width=4in]{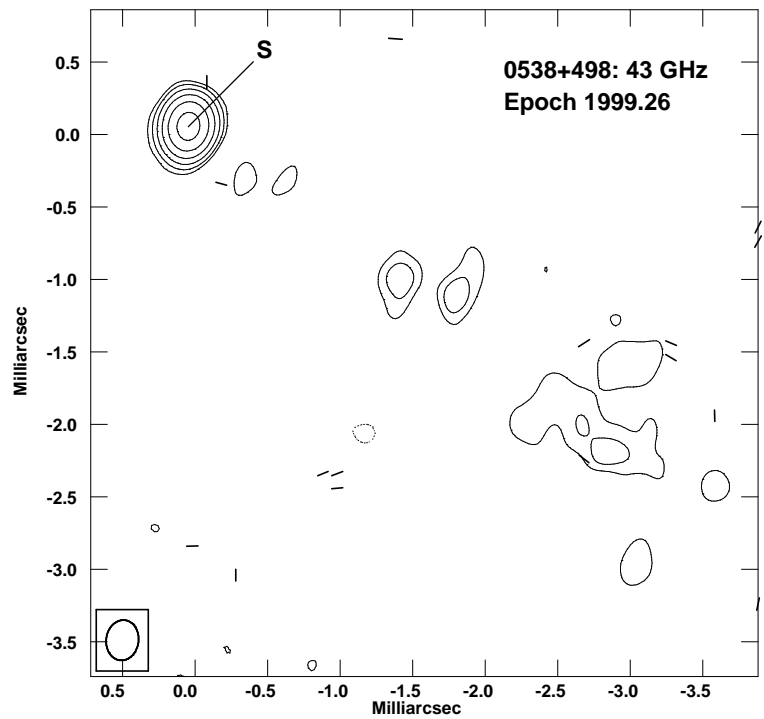}
\caption{\label{map7} VLBA total intensity image of 0538+498 (3C~147) at 43 GHz,
 with electric vectors superimposed.  }
\end{figure*}

\begin{figure*}
\centering
\includegraphics[width=4in]{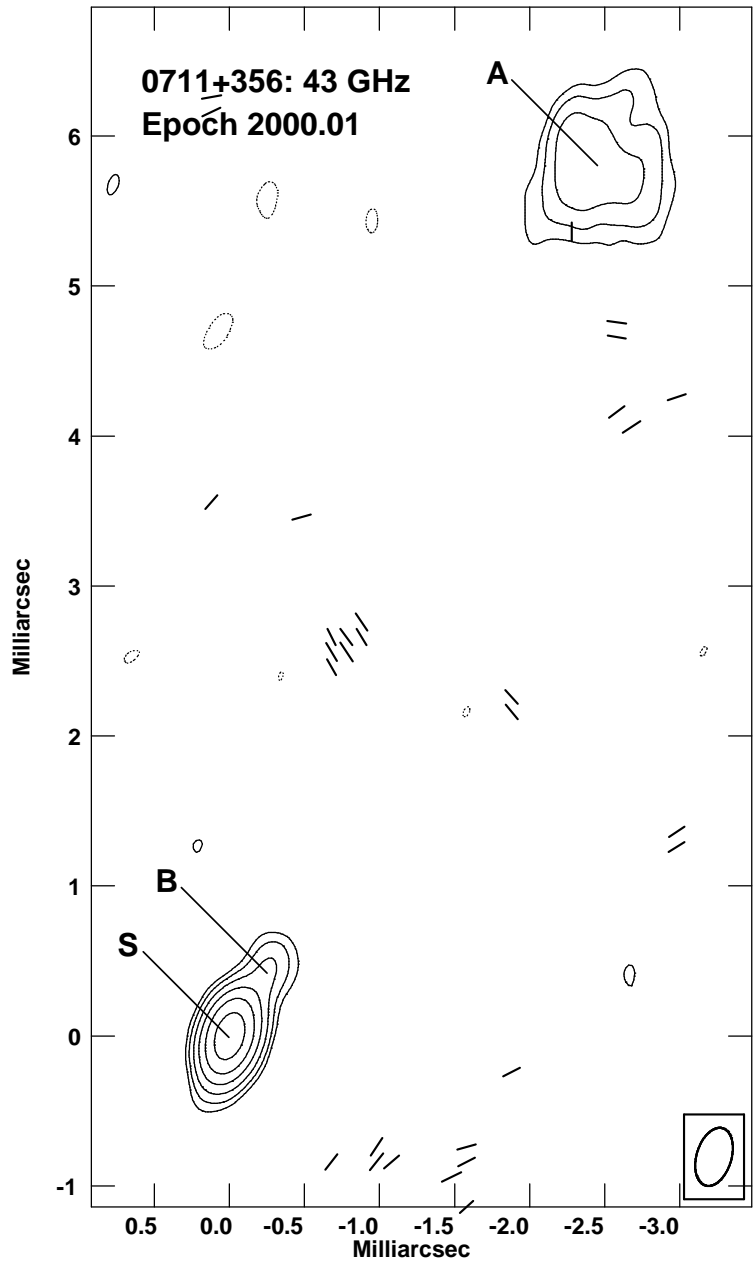}
\caption{\label{map8} VLBA total intensity image of 0711+356 (OI~318) at 43 GHz,
 with electric vectors superimposed.  }
\end{figure*}

\begin{figure*}
\centering
\includegraphics[angle = -90,width=4in]{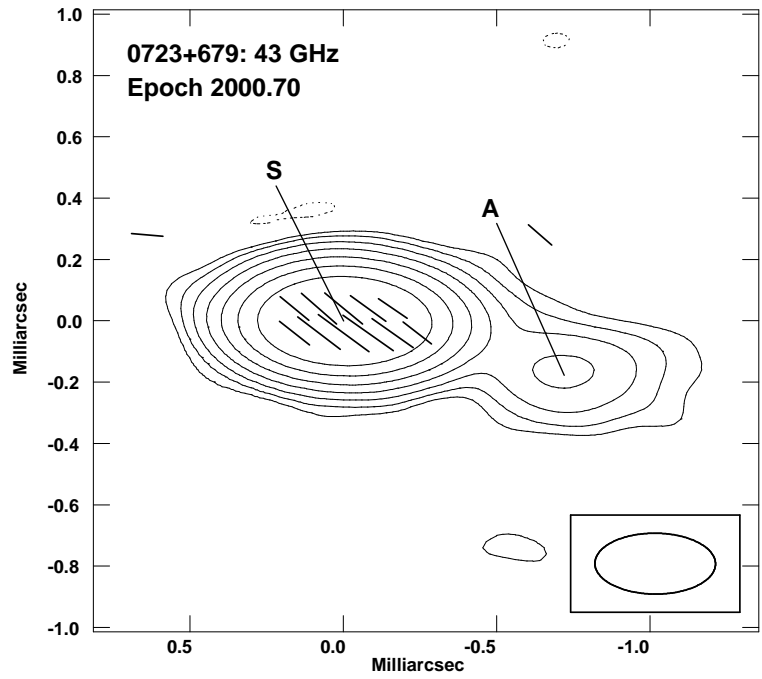}
\caption{\label{map9} VLBA total intensity image of 0723+679 (3C~179) at 43 GHz, 
with electric vectors superimposed. }
\end{figure*}

\begin{figure*}
\centering
\includegraphics[angle = -90,width=4in]{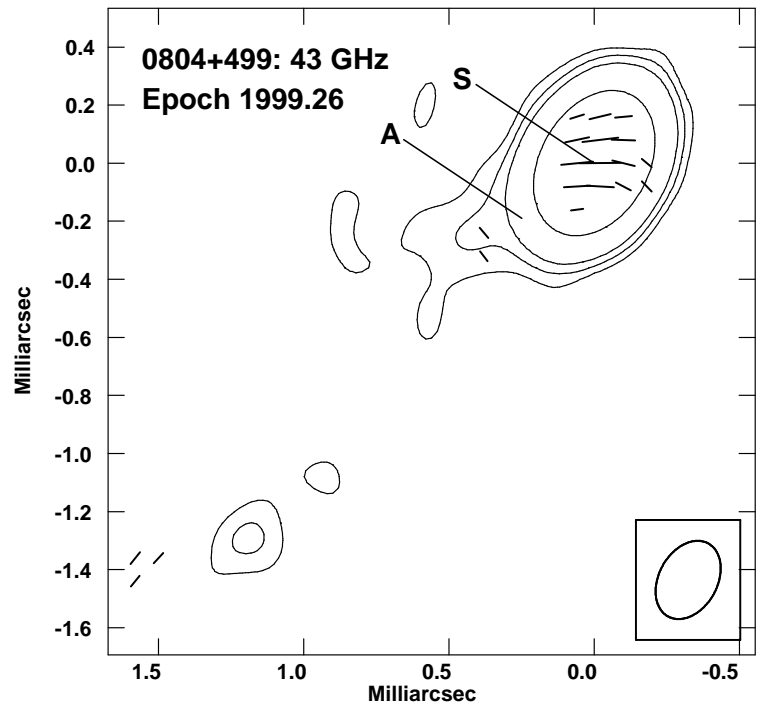}
\caption{\label{map10} VLBA total intensity image of 0804+499 (OJ~508) at 43 GHz,
 with electric vectors superimposed.  }
\end{figure*}

\begin{figure*}
\centering
\includegraphics[angle = 0,width=4.5in]{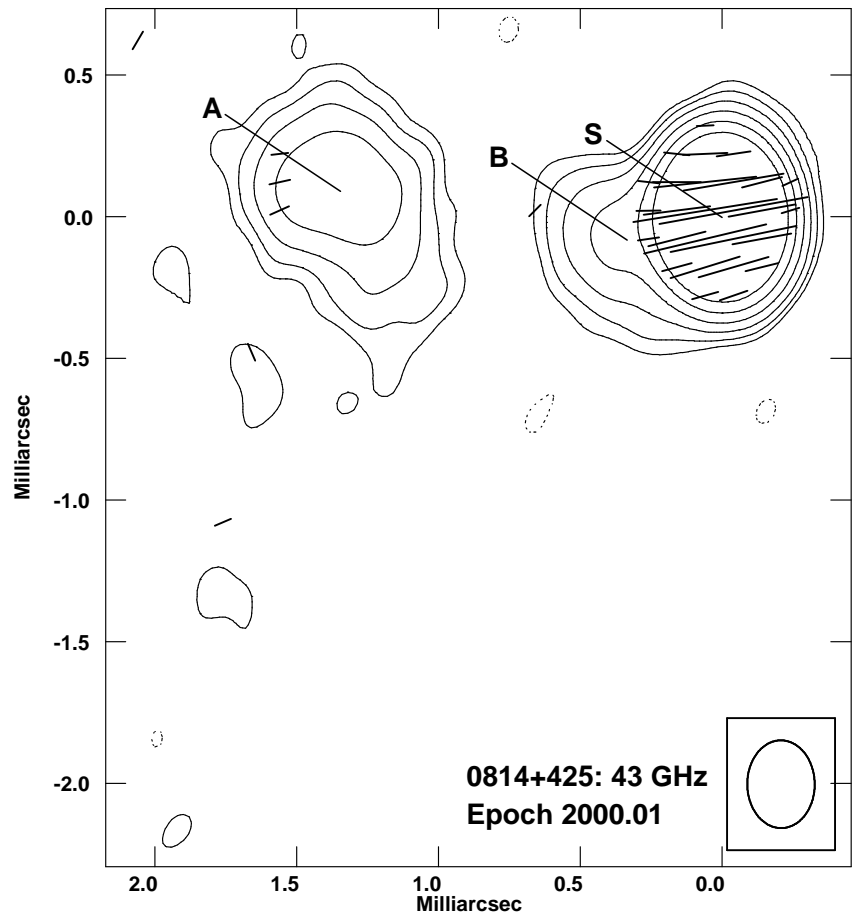}
\caption{\label{map11} VLBA total intensity image of 0814+425 (OJ~425) at 43 GHz, 
with electric vectors superimposed. }
\end{figure*}

\begin{figure*}
\centering
\includegraphics[angle = 0,width=4.5in]{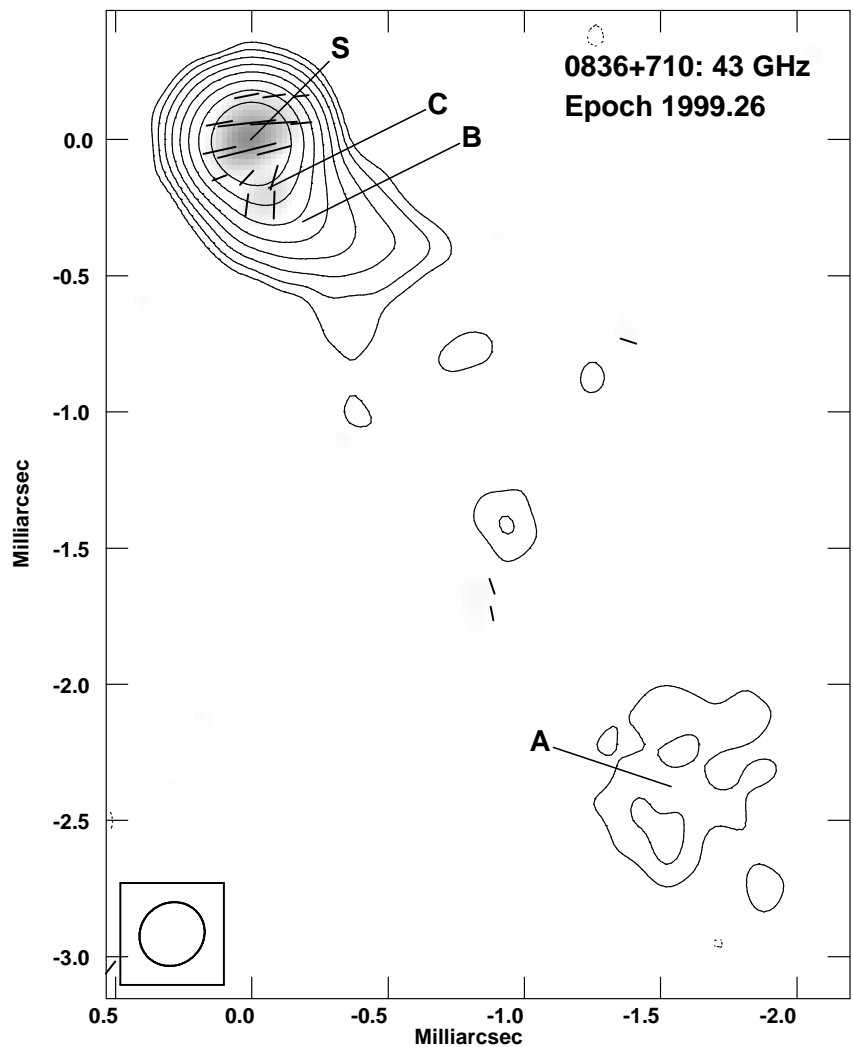}
\caption{\label{map12} VLBA total intensity image of 0836+710 (4C~71.07) at 43 GHz, 
with electric vectors superimposed. The gray scale
indicates linearly polarized intensity. }
\end{figure*}

\begin{figure*}
\centering
\includegraphics[angle = 0,width=4in]{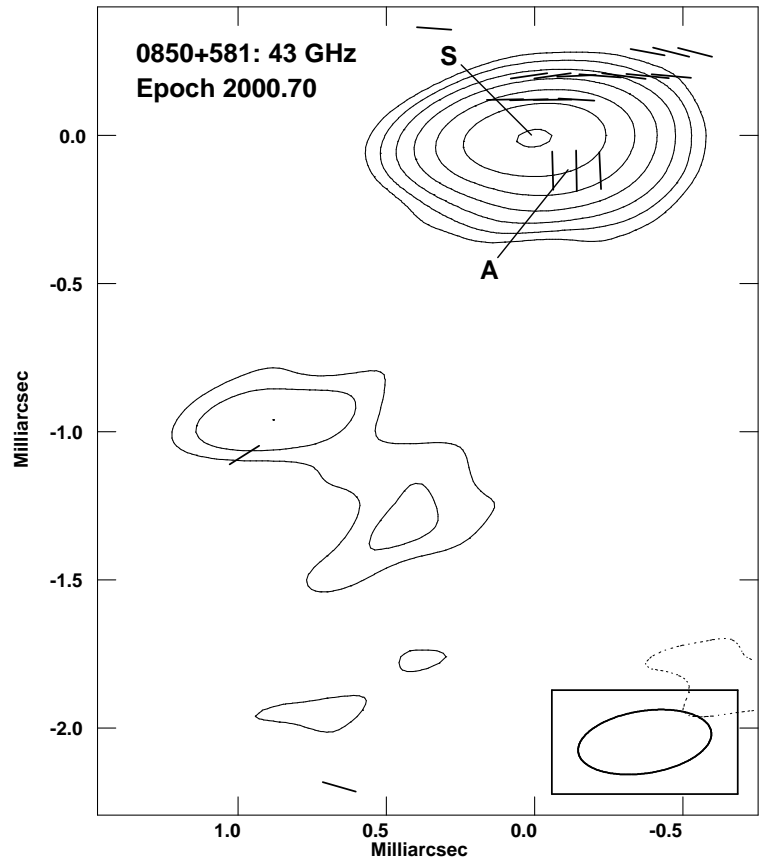}
\caption{\label{map13} VLBA total intensity image of 0850+581 (4C~58.17) at 43 GHz, 
with electric vectors superimposed. }
\end{figure*}

\begin{figure*}
\centering
\includegraphics[angle = 0,width=4in]{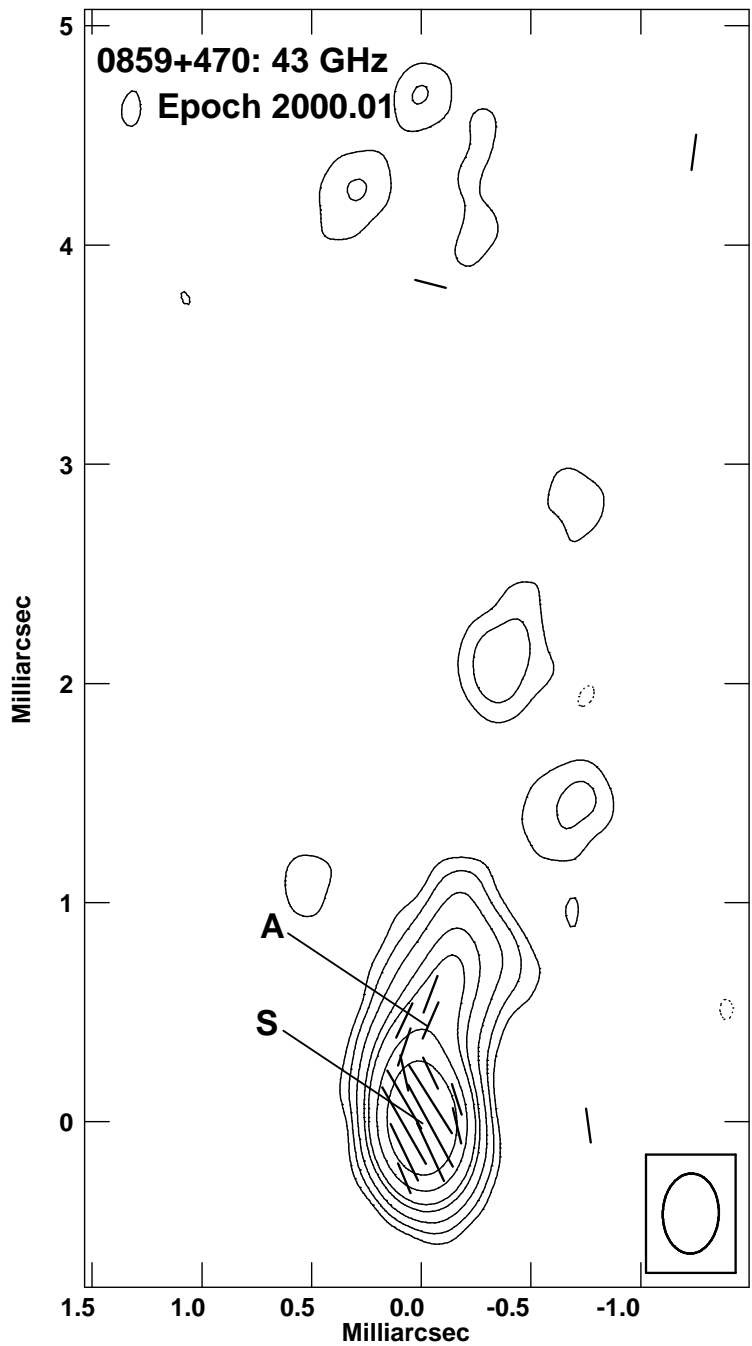}
\caption{\label{map14} VLBA total intensity image of 0859+470 (4C~47.29) at 43 GHz, 
with electric vectors superimposed. }
\end{figure*}

\begin{figure*}
\centering
\includegraphics[angle = 0,width=4in]{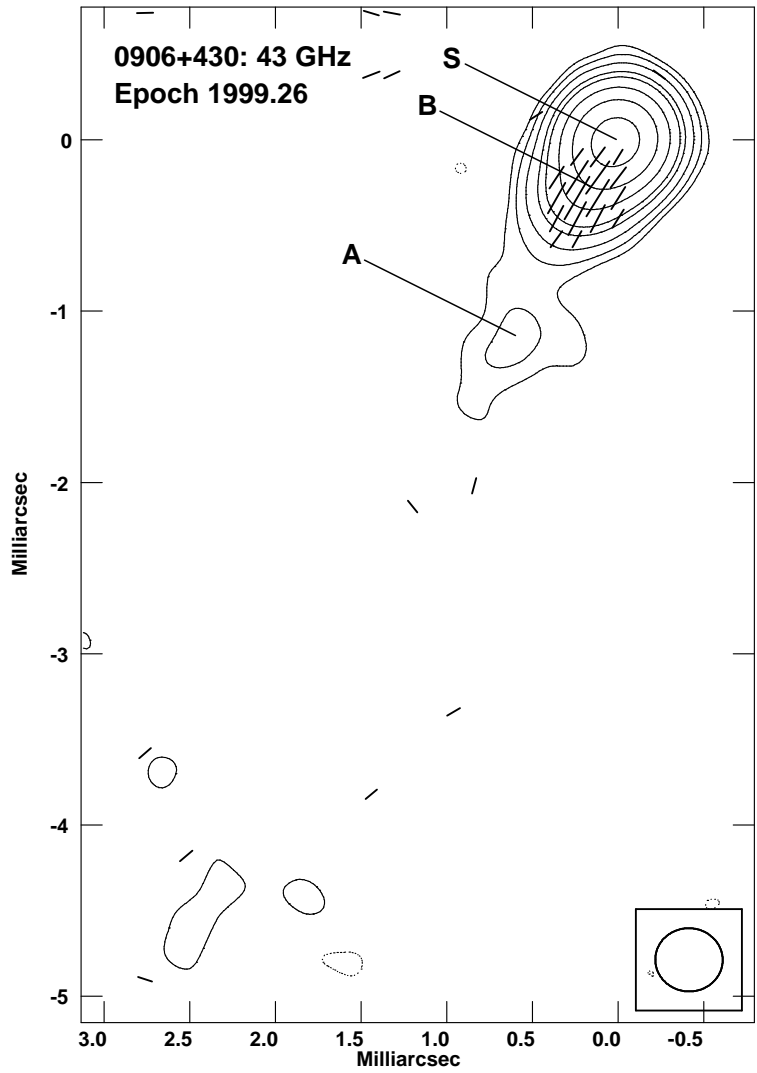}
\caption{\label{map15} VLBA total intensity image of 0906+430 (3C~216) at 43 GHz, 
with electric vectors superimposed. }
\end{figure*}

\begin{figure*}
\centering
\includegraphics[angle = -90,width=5in]{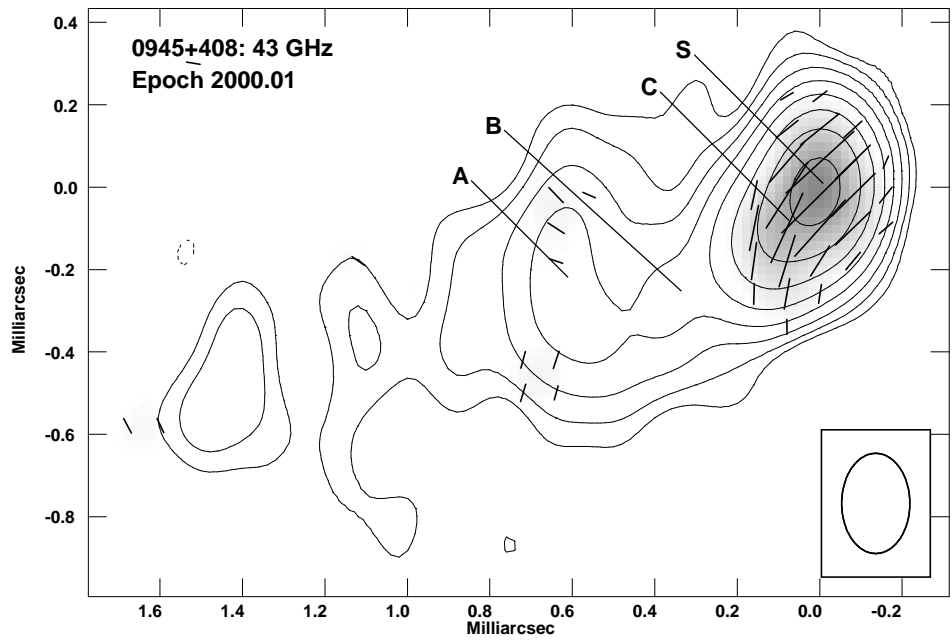}
\caption{\label{map16} VLBA total intensity image of 0945+408 (4C~40.24) (3C~179) at 43 GHz, 
with electric vectors superimposed. The gray scale
indicates linearly polarized intensity. }
\end{figure*}

\begin{figure*}
\centering
\includegraphics[angle = 0,width=4in]{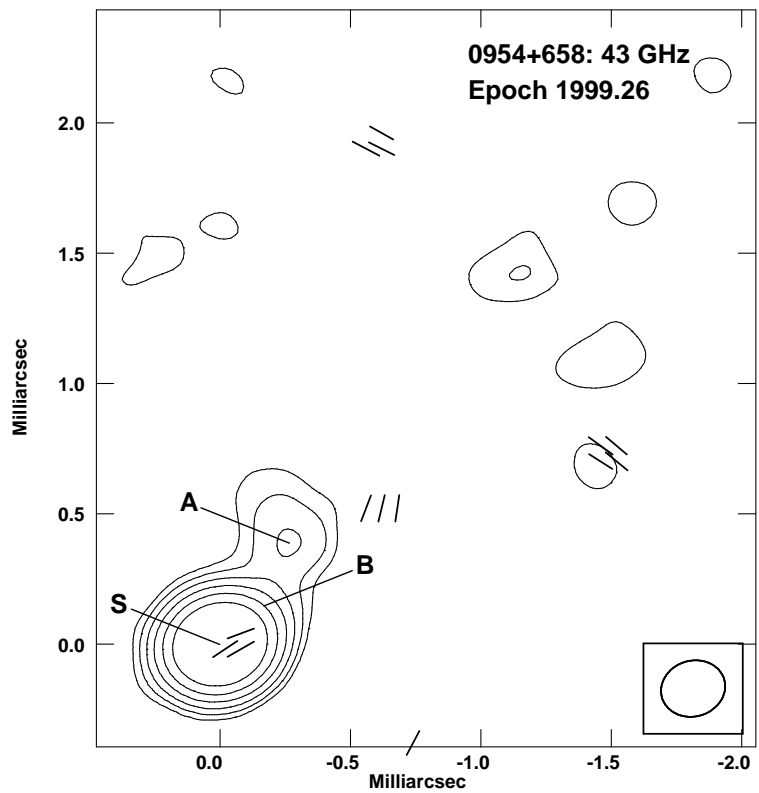}
\caption{\label{map17} VLBA total intensity image of 0954+658  at 43 GHz, 
with electric vectors superimposed. }
\end{figure*}

\begin{figure*}
\centering
\includegraphics[angle = -90,width=5.5in]{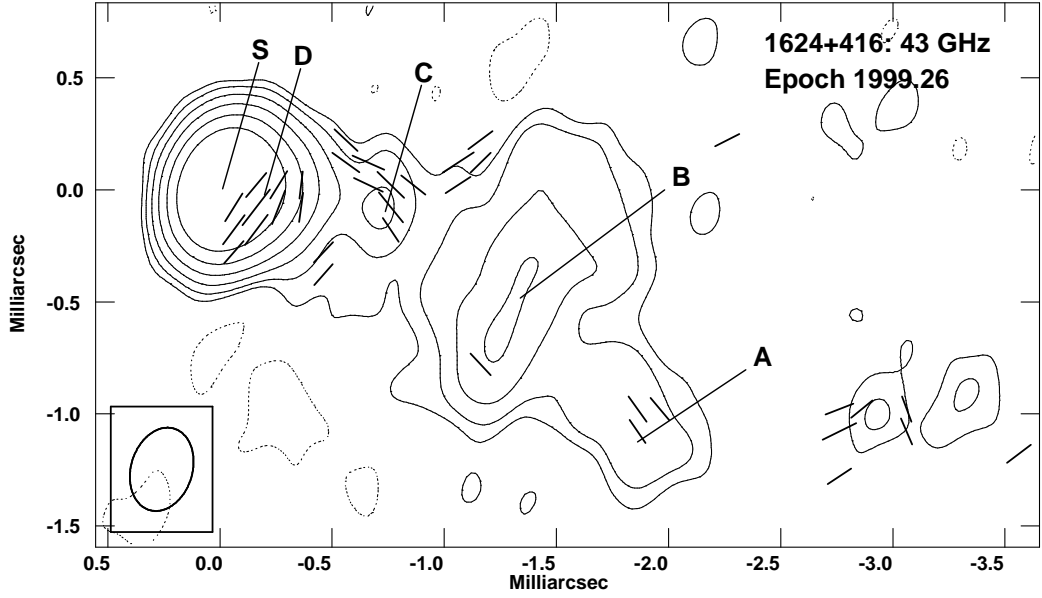}
\caption{\label{map18} VLBA total intensity image of 1624+416 (4C~41.32) at 43 GHz, 
with electric vectors superimposed.  }
\end{figure*}

\begin{figure*}
\centering
\includegraphics[angle = 0,width=3in]{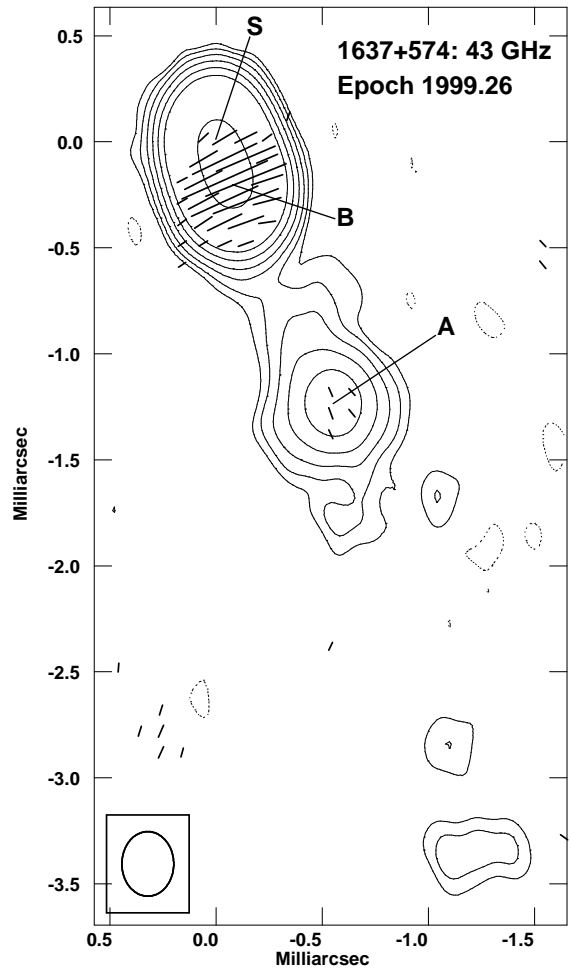}
\caption{\label{map19} VLBA total intensity image of 1637+574 (OS~562) at 43 GHz, 
with electric vectors superimposed.  }
\end{figure*}

\begin{figure*}
\centering
\includegraphics[angle = -90,width=5.5in]{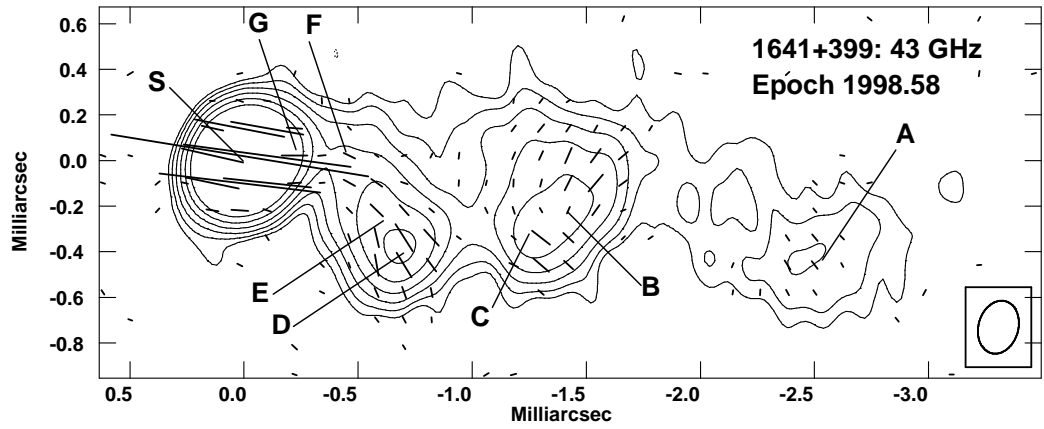}
\caption{\label{map20} VLBA total intensity image of 1641+399 (3C~345) at 43 GHz, 
with electric vectors superimposed. }
\end{figure*}

\begin{figure*}
\centering
\includegraphics[angle = 0,width=5in]{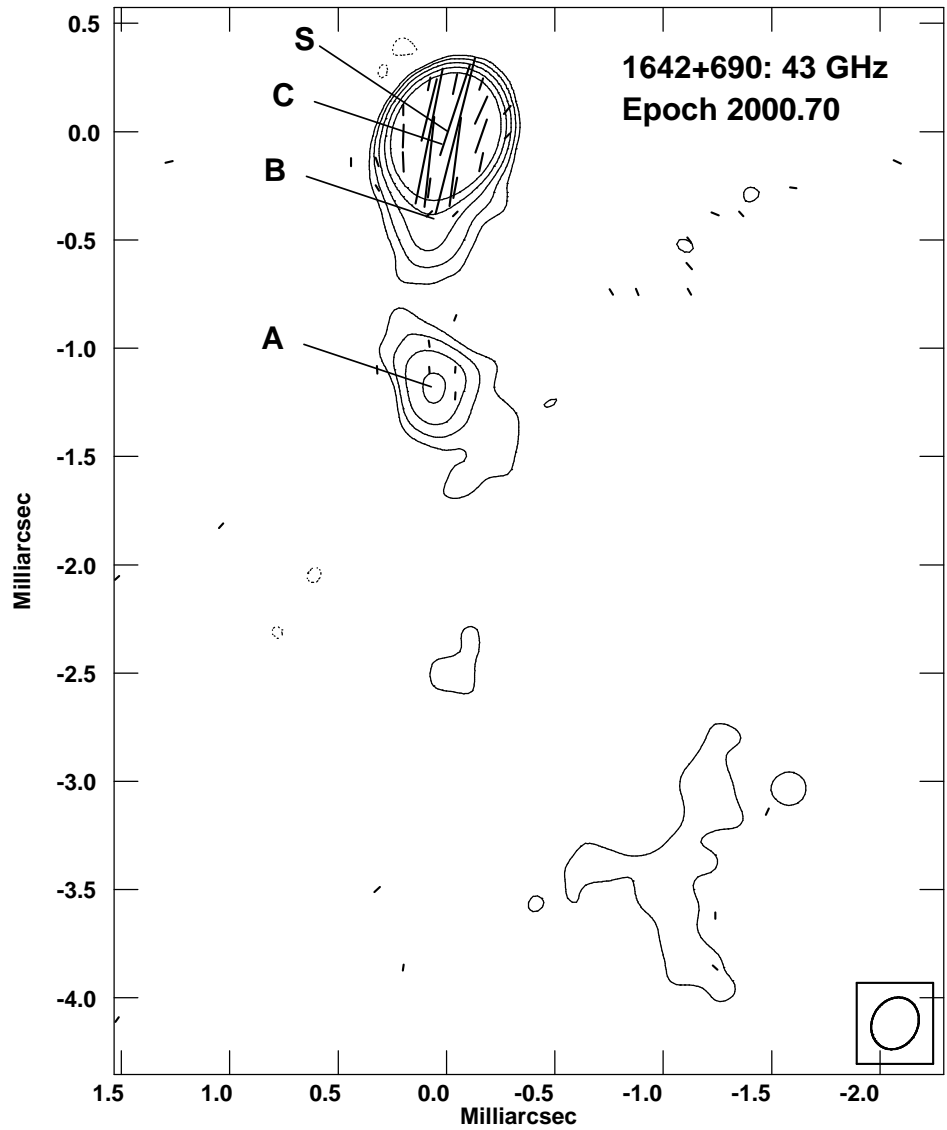}
\caption{\label{map21} VLBA total intensity image of 1642+690 (4C~69.21) at 43 GHz, 
with electric vectors superimposed. }
\end{figure*}

\begin{figure*}
\centering
\includegraphics[angle = -90,width=5.5in]{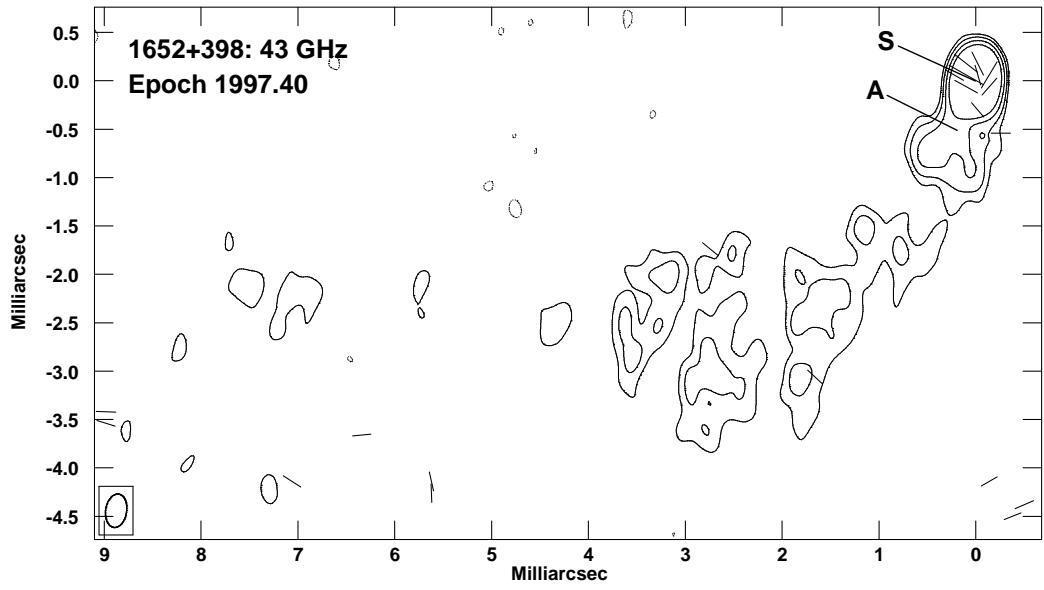}
\caption{\label{map22} VLBA total intensity image of 1652+398 (MK~501) at 43 GHz, 
with electric vectors superimposed.  }
\end{figure*}

\begin{figure*}
\centering
\includegraphics[angle = 0,width=4in]{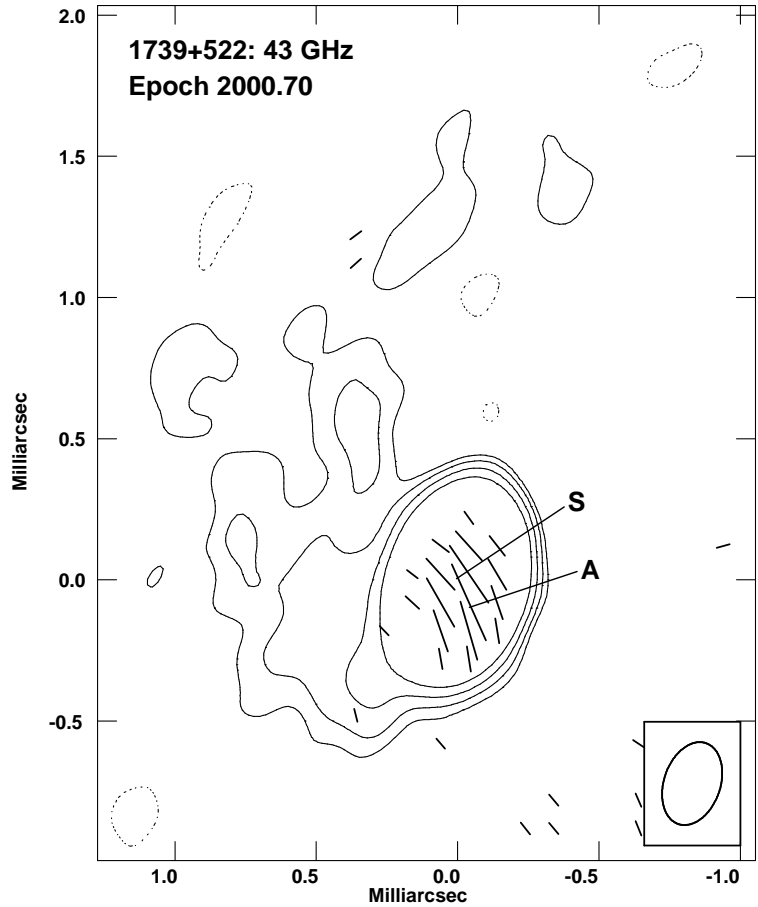}
\caption{\label{map23} VLBA total intensity image of 1739+522 (4C~51.37) at 43 GHz, 
with electric vectors superimposed. }
\end{figure*}

\begin{figure*}
\centering
\includegraphics[angle = -90,width=5in]{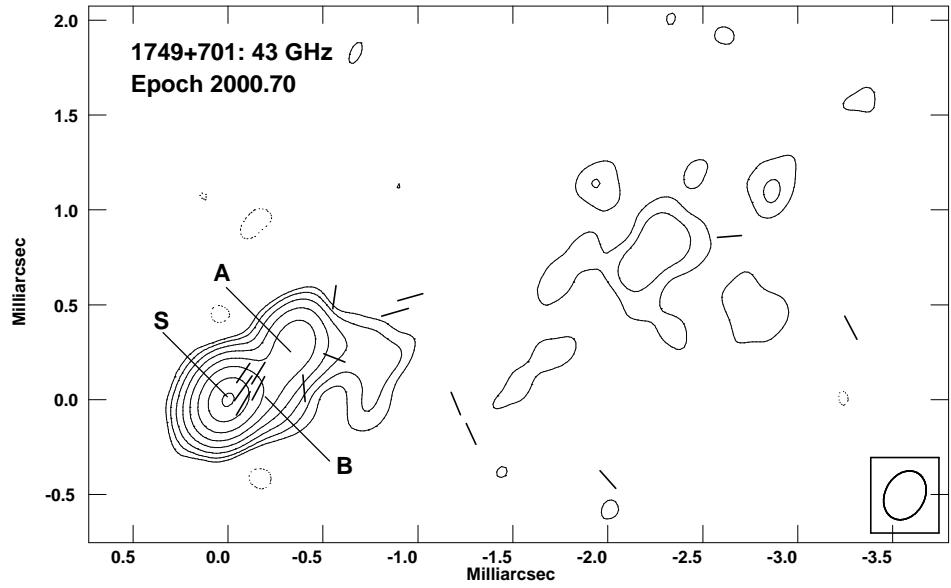}
\caption{\label{map24} VLBA total intensity image of 1749+701  at 43 GHz, 
with electric vectors superimposed. }
\end{figure*}

\begin{figure*}
\centering
\includegraphics[angle = -90,width=5in]{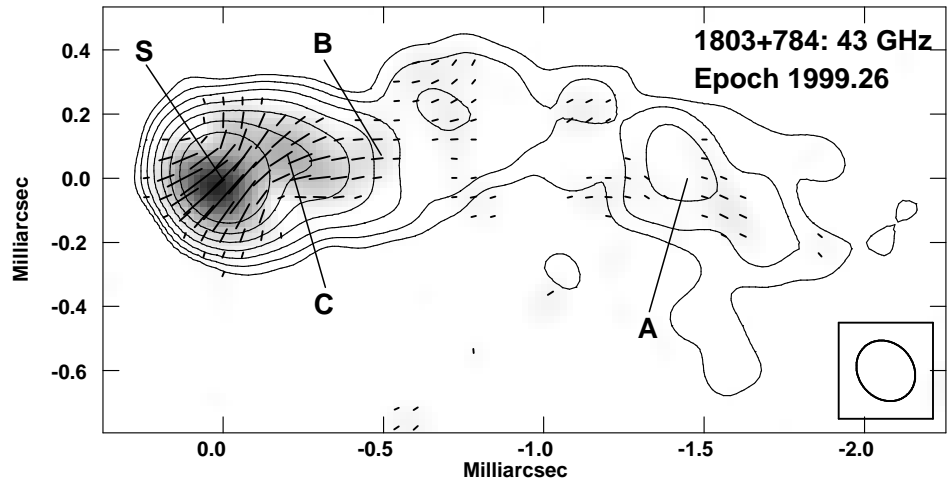}
\caption{\label{map25} VLBA total intensity image of 1803+784  at 43 GHz, 
with electric vectors superimposed. The gray scale
indicates linearly polarized intensity. }
\end{figure*}

\begin{figure*}
\centering
\includegraphics[angle = -90,width=5in]{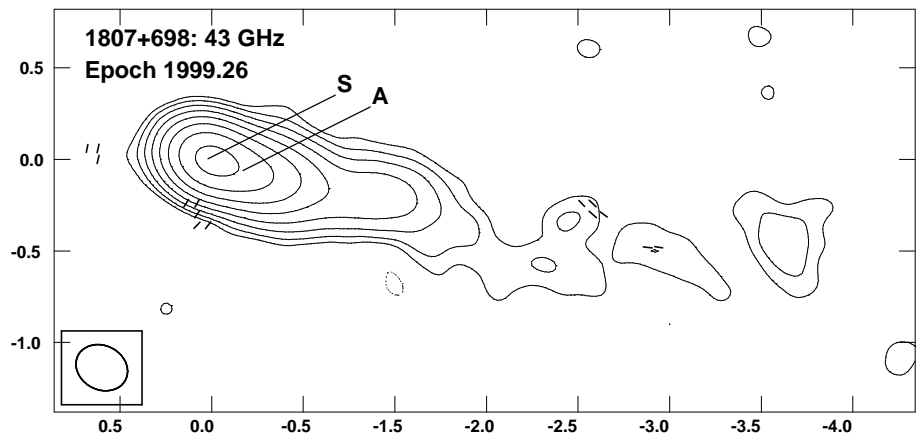}
\caption{\label{map26} VLBA total intensity image of 1807+698 (3C~371) at 43 GHz, 
with electric vectors superimposed. }
\end{figure*}

\begin{figure*}
\centering
\includegraphics[angle = 0,width=4.5in]{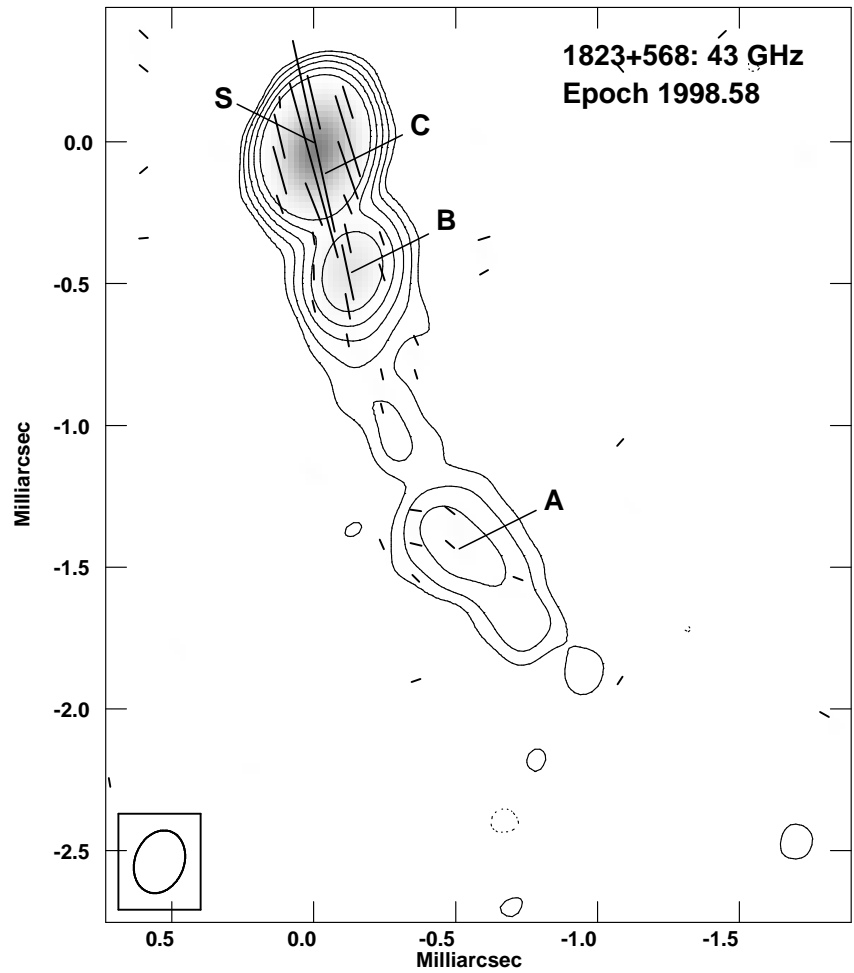}
\caption{\label{map27} VLBA total intensity image of 1823+568 (4C~56.27) at 43 GHz, 
with electric vectors superimposed. The gray scale
indicates linearly polarized intensity. }
\end{figure*}

\begin{figure*}
\centering
\includegraphics[angle = 0, width = 6.5in]{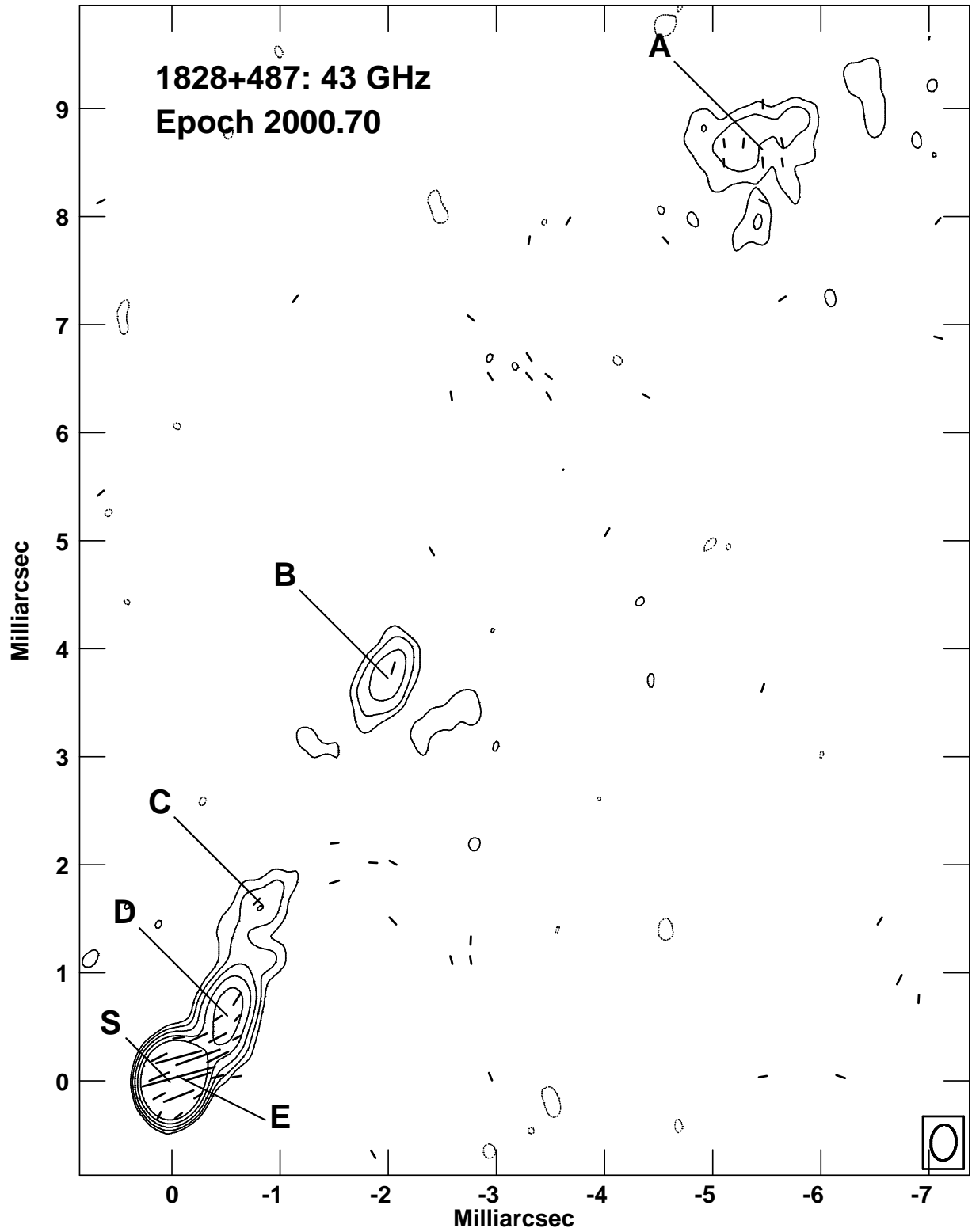}
\caption{\label{map28} VLBA total intensity image of 1828+487 (3C~380) at 43 GHz, 
with electric vectors superimposed. }
\end{figure*}

\begin{figure*}
\centering
\includegraphics[angle = -90,width=5.5in]{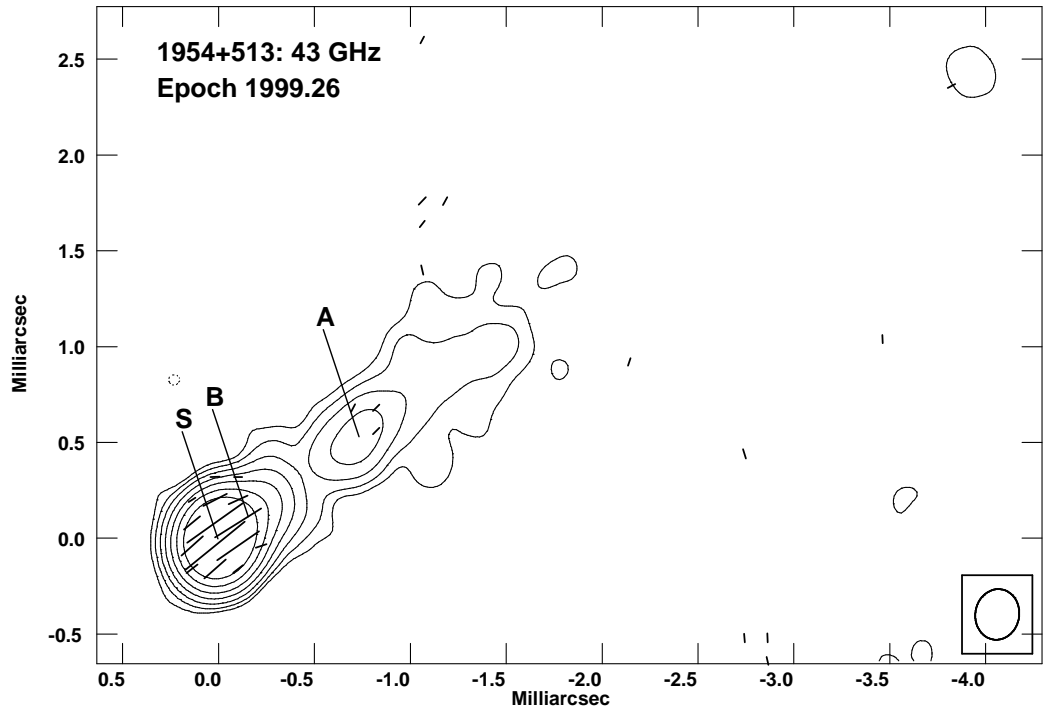}
\caption{\label{map29} VLBA total intensity image of 1954+513 (OV~591) at 43 GHz, 
with electric vectors superimposed. }
\end{figure*}

\begin{figure*}
\centering
\includegraphics[angle = 0,width=4.5in]{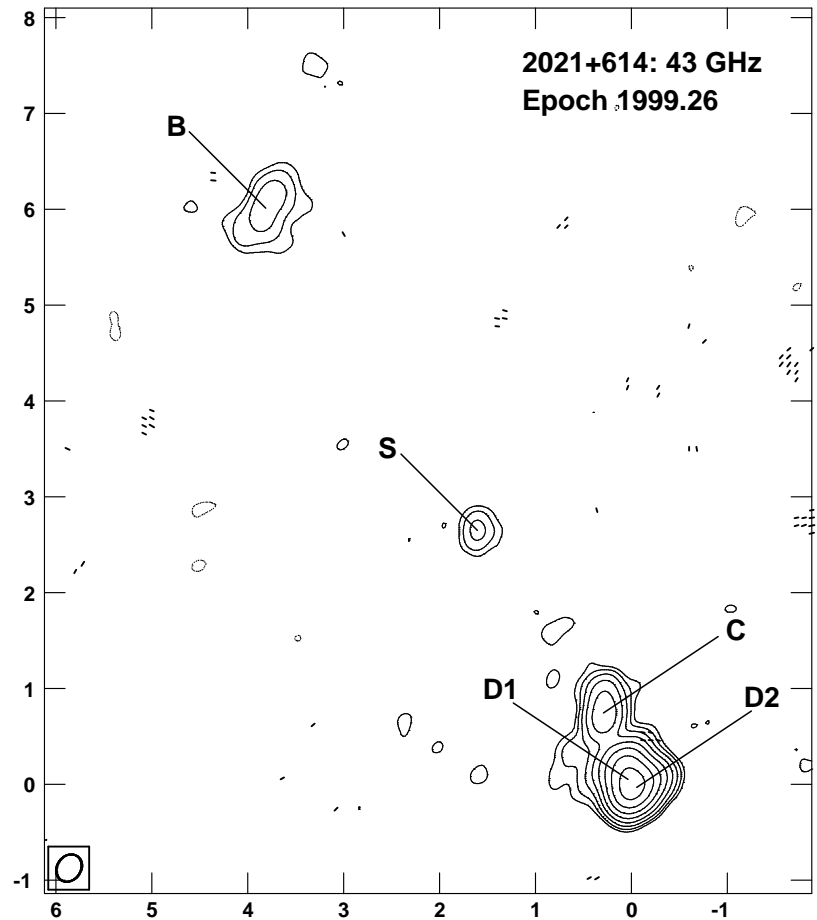}
\caption{\label{map30} VLBA total intensity image of 2021+614 (OW~637) at 43 GHz, 
with electric vectors superimposed. }
\end{figure*}

\begin{figure*}
\centering
\includegraphics[angle = 0,width=5in]{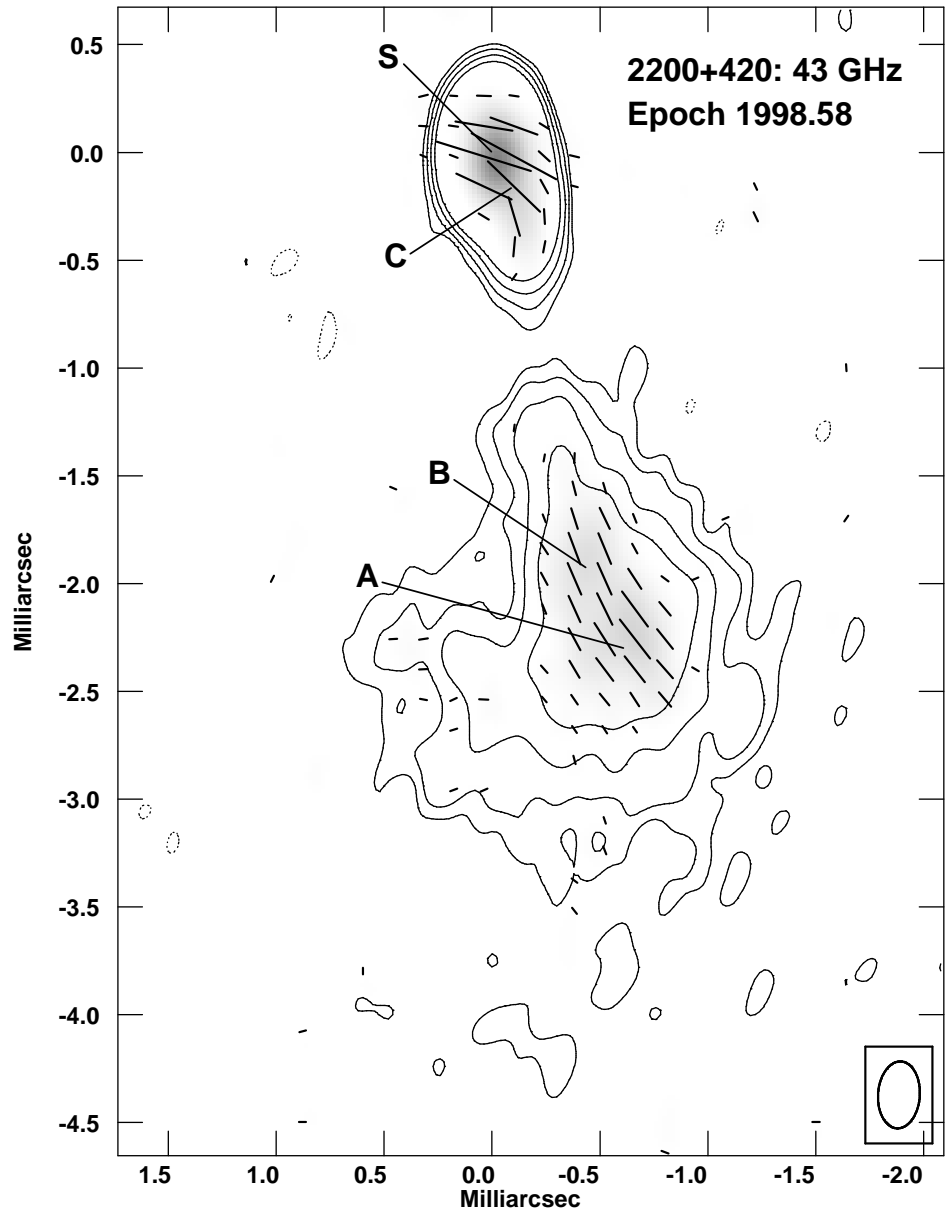}
\caption{\label{map31} VLBA total intensity image of 2200+420 (BL~Lac) at 43 GHz, 
with electric vectors superimposed. The gray scale
indicates linearly polarized intensity. }
\end{figure*}

\begin{figure*}
\centering
\includegraphics[angle = -90,width=5in]{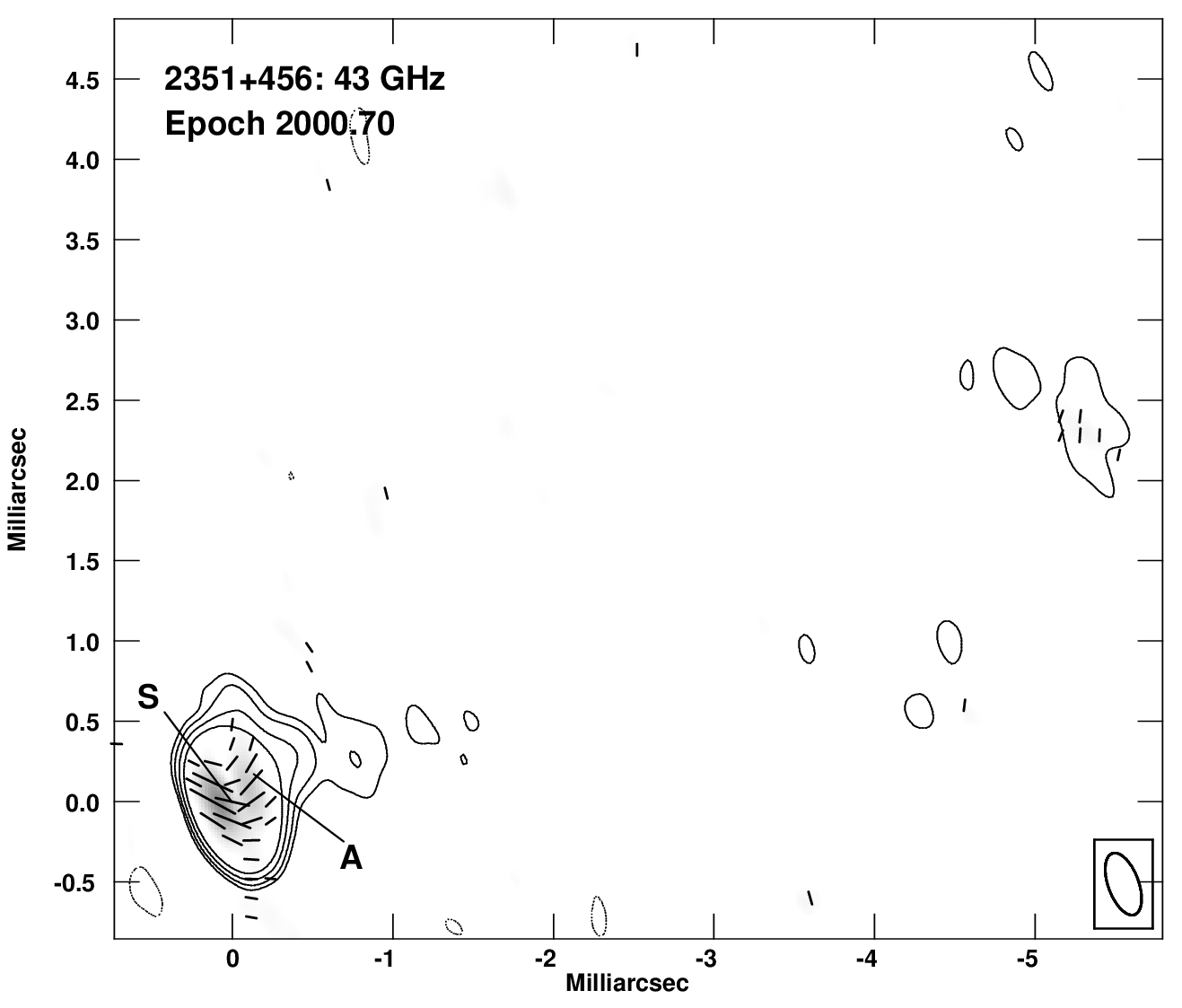}
\caption{\label{map32} VLBA total intensity image of 2351+456 (4C~45.51) at 43 GHz, 
with electric vectors superimposed. The gray scale
indicates linearly polarized intensity. }
\end{figure*}

We used the Caltech Difmap package \citep{SPT94} to make Stokes $I$,
$Q$, and $U$ images of our sources, which we then imported into AIPS
to make our final contour images, shown in
Figures~\ref{map1}-\ref{map32}. The images of 0923+392, 1633+382, and
1928+738 are presented in \cite{LS00}. In the interest of completeness
we include unpublished images of 1652+398 from A. P. Marscher (private
communication) and those of 1641+399, 1823+487, and 2200+420 from
S. G. Jorstad (in preparation).  The parameters for all of the images
are summarized in Table~\ref{mapstats}.

\clearpage

We performed elliptical Gaussian model fits to each source in the {\it
(u,v)} plane using the task ``modelfit'' in Difmap. In cases where
elliptical Gaussians provided a poor fit to the visibilities, we used
point ($\delta$-function) components. The results of these fits are given in
Tables~\ref{coreprops} and \ref{jetprops}, and are intended as a
general guideline for interpreting the source structure.  We estimate
the given positions of strong, isolated components to be accurate to
within a quarter of a beam width, and their measured flux densities to
an accuracy of $\lesssim 10\%$. Our fits are less reliable for weak
components and those located in regions of diffuse emission.

At the central position of each component we measured the electric
vector position angle (EVPA) and the percentage linear polarization
$m = 100 \times P/I$, where $P=({Q^2+U^2})^{1/2}$, and $Q$, $U$, and
$I$ are Stokes flux densities. We note that in some cases the peak of
polarized emission is offset from that of the $I$ emission, so that
the values of $m$ given in Tables~\ref{coreprops} and
\ref{jetprops} may not represent the maximum percentage polarization
associated with a particular component.  For those components with a
peak $P$ flux density smaller than five times the rms noise level in
the P image, we quote an upper limit on $m$. We make no corrections
for Ricean bias in the other components, since these are all negligible.

\section{\label{discussion} Discussion: Correlations among source properties}
The Pearson-Readhead sample has been extensively studied at a variety
of wavelengths and resolution levels, and has one of the largest
collections of published data of any radio sample.  Many of these data
were tabulated by \cite{LTP01} and are summarized here in
Tables~\ref{genprops} and \ref{quantities}. 

We have extended the multi-dimensional analysis of \cite{LTP01} to
look for trends between our measured polarization properties and
various source quantities from the literature using Kendall's tau
tests.  For those source properties that contain measured upper or
lower limits (i.e., censored data), we used a version of Kendall's tau
test from the Astronomical Survival Analysis (ASURV) package
\citep{ASURV}. Many of the observed properties in flux-limited AGN
samples (e.g., luminosity) have a strong redshift dependence, due to
the lower flux cutoff and the steepness of the parent luminosity
function \citep{Pad92}. We therefore calculated an additional partial
correlation coefficient removing the effects of redshift (see
\citealt{LTP01}) for each pair of variables. In the case of variables
involving censored data, we used the algorithm of \cite{AS96}.  Given
the large number of tests performed, we rejected any correlations with
a confidence level less than $\sim 98\%$. We also rejected any
instances in which one or more outlying points artificially raised the
correlation significance above this level. We give a list of
significant correlations and their statistical confidence levels in
Table~\ref{corr_results}.

We also performed a series of two-sample tests on the data to look for
differences in properties among different source classes, such as
quasars versus BL Lacertae objects. For the non-censored variables, we
used a standard Kolmogorov-Smirnov test, while for the censored ones,
we used Gehan's generalized Wilcoxon two-sample test \citep{G65} from
the ASURV package. We again used a $98\%$ confidence level to
determine whether any differences were significant. The results of
these tests can be found in Table~\ref{ks_results}, and will be
discussed in \S~5. Here we discuss the individual correlations we have
found using Kendall's tau tests for the FS-PR sample.

\subsection{Redshift}
We find that all of the luminosity-related quantities of the FS-PR
sample have a strong dependence on redshift due to the flux limited
nature of the sample. This includes the variability brightness
temperature (99.97\% confidence;
Table~\ref{corr_results}), which  \cite{LV99} calculated for 20
FS-PR objects by using the observed time scale of a well-defined flare
at 37 GHz to estimate the size of the emitting region. The strong
correlation with redshift is likely due to the fact that the
variability brightness temperature is directly proportional to the
observed flux density (and hence luminosity) of the flare.

\subsection{\label{lum_lum} Luminosity properties}

At 43 GHz, nearly all of the FS-PR sources have radio morphologies
that are dominated by a strong core component located near the base
of an extended jet. Multi-frequency studies have shown that these
cores have flat or inverted spectra, and likely represent the region
where the jet becomes optically thick. The core in lower frequency
VLBI images often turns out to be a blend of two or more components
due to insufficient spatial resolution \citep{LS00}. For the majority
of the FS-PR sources our 43 GHz observations represent the highest
resolution images obtained to date, and are less subject to possible
blending effects.

We have searched for trends between the luminosity of the VLBI core at
43 GHz and luminosities at other wavelengths using the partial
correlation coefficients described in \S~4. We find that both the
total VLBI luminosity of the source and the luminosity of the VLBI
core are well-correlated with a) the VLBI core luminosity at 5 GHz and
b) the x-ray luminosity at 1 keV ($L_{xray}$). \cite{BMM99} found a
correlation between 22 GHz VLBI core luminosity and $L_{xray}$ using a
different flat-spectrum AGN sample, and attributed it to a common synchrotron origin
for these spectral components.  We also detect a correlation between
43 GHz VLBI core luminosity and optical luminosity (see
Table~\ref{corr_results}).  Taken together, these findings suggest
that there is a well-defined synchrotron component in these AGNs whose
emission extends from millimeter wavelengths through optical to x-ray
energies.

\subsection{Core dominance}
The highly core-dominated nature of the FS-PR sample is a reflection
of its selection criteria and the dimming of steep-spectrum jet
emission at high frequency. In order to investigate possible
correlations between core dominance and other properties, we
define a parameter $R$ (column 7 of Table~\ref{coreprops})
as the flux density of the core divided by the remaining (jet) flux
density in the VLBI image. We correct the core dominance to the
quasar rest frame by assuming typical spectral indices of 0 and $-0.8$
for the core and jet emission, respectively. The $R$ values range from
0.02 to $\sim20$, with a median of 0.75. This suggests that most
bright, flat-spectrum AGNs should be excellent targets for next
generation high frequency space-VLBI missions, since they are likely
compact enough to have sufficient flux density for fringe detection on
extremely long baselines.

\begin{figure*}
\plotone{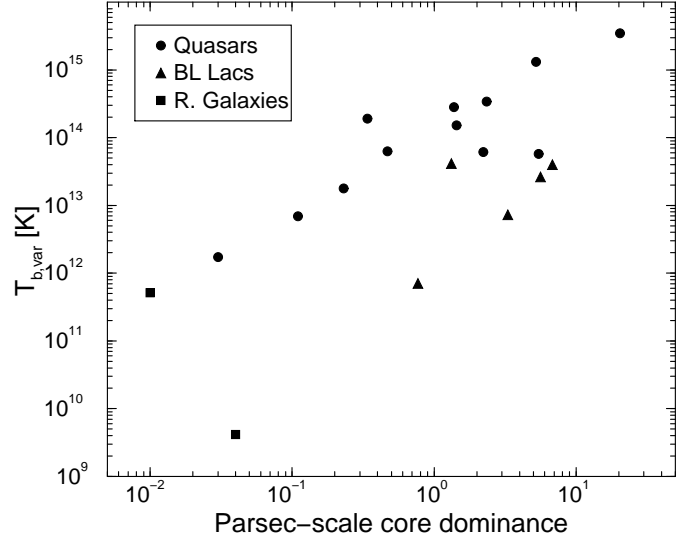}
\caption{\label{r__Tb_var} Variability brightness temperature at 37
GHz versus parsec-scale core dominance ($R$) at 43 GHz.}
\end{figure*}

Our $R$ parameter is somewhat analogous to the core dominance
parameter $F_c$ of \cite{PR88}, which they defined as the ratio of
VLBI core to total (single dish) flux density of the source at 5
GHz. The $F_c$ parameter is considered to be a good indicator of
Doppler beaming, which causes a dramatic boosting of the jet and core
fluxes in relativistic jets viewed nearly end-on. The extended radio
lobes of AGNs are not as affected by beaming, since their bulk motions
are non-relativistic. Therefore, the ratio of core-to-total flux
density will increase as an AGN is viewed closer to the jet axis. This
effect appears to be responsible for the well-known correlations
between core dominance and a) optical polarization \citep{ILT91} and
b) 5 GHz radio variability \citep{AAH92} seen in the full PR
sample. The latter two quantities are correlated with our $R$
parameter at the 99.89\% and 99.45\% confidence level,
respectively. We find a good correlation (98.93\% confidence;
Fig.~\ref{r__Tb_var}) between $R$ and another quantity known to be
influenced by beaming, namely the variability brightness temperature
\citep{LV99}.

The number of correlations between our $R$ parameter and other known beaming
indicators is somewhat surprising, since there is strong evidence
(e.g., from rapid flux variability and superluminal motions) that {\it
both} the core and jet emission on VLBI scales are beamed.  Nevertheless we find
evidence that the ratio of core-to-jet flux is higher in the more
highly beamed sources. This runs contrary to the predictions of the
standard jet model, which suggests that the core emission should be
{\it less} boosted than that of the jet. Since the jet emission is
dominated by components with finite lifetimes, its emission will be
boosted by a factor $\delta^{3-\alpha}$, where $\delta$ is the Doppler
factor of the bulk flow. The core on the other hand will only be
boosted by a factor $\delta^{2-\alpha}$.

These apparent contradictions can be resolved by considering several
other effects that can influence the $R$ values of beamed sources.
First, we expect the high-energy electrons that are accelerated in the
cores to experience inverse-Compton losses from scattering off of
external photons. These losses will be larger for highly beamed
sources, due to blue-shifting of the external photons in the rest
frame of the jet (e.g., \citealt{DS93}). According to the shock-in-jet model of
\cite{MG85}, the lifetimes of components that evolve adiabatically as
they move down the jet will be shortened accordingly, due to the
steepening of their electron energy distribution and the shifting of
the high-energy electron cutoff into the radio region. This will tend
to reduce the jet flux relative to the core and increase $R$. Second,
highly beamed sources will tend to have faster component motions,
which will cause the jet components to fade out more rapidly, thereby
reducing the total jet flux and lowering the number of components per
unit length in the jet. Third, it is possible that the bulk speed of
the flow is higher in the cores than further down the jet, which would
lead to a higher amount of beaming in the core. Finally, the standard
jet model predictions are based on the assumption that the core
emission originates from a continuous jet, however, our VLBI ``cores''
may include emission from unresolved components very close to the base
of the jet which have finite lifetimes.

\subsection{Fractional polarization of core components}
\begin{figure*}
\plotone{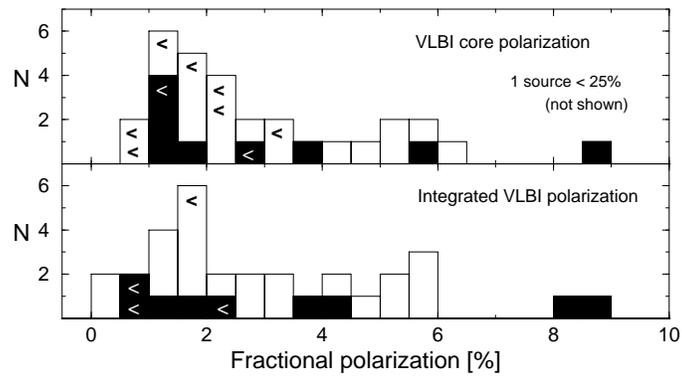}
\caption{\label{m_plots} Top panel: fractional 
polarization distribution of FS-PR core components at 43 GHz.  Bottom
panel: integrated 43 GHz VLBI fractional polarization distribution of FS-PR
sources. The shaded boxes in both panels represent BL Lac objects. }
\end{figure*}

The fractional polarizations of the VLBI core components in our sample
range up to 9\%, with most being weakly polarized ($\lesssim 2\%$).
The integrated VLBI fractional polarization (defined as the ratio of
$(Q^2 + U^2)^{1/2}$ to I, where Q, U, and I represent the cleaned Stokes
fluxes in the VLBI images) has a similar distribution to that of the
cores (Figure~\ref{m_plots}). This is simply a reflection of the
dominance of the core component and relatively weak contribution of
the steep spectrum jet to the polarized flux at 43 GHz.  Indeed, we
find that the core and integrated VLBI polarization levels are
correlated at the 99.995\% confidence level.

\begin{figure*}
\plotone{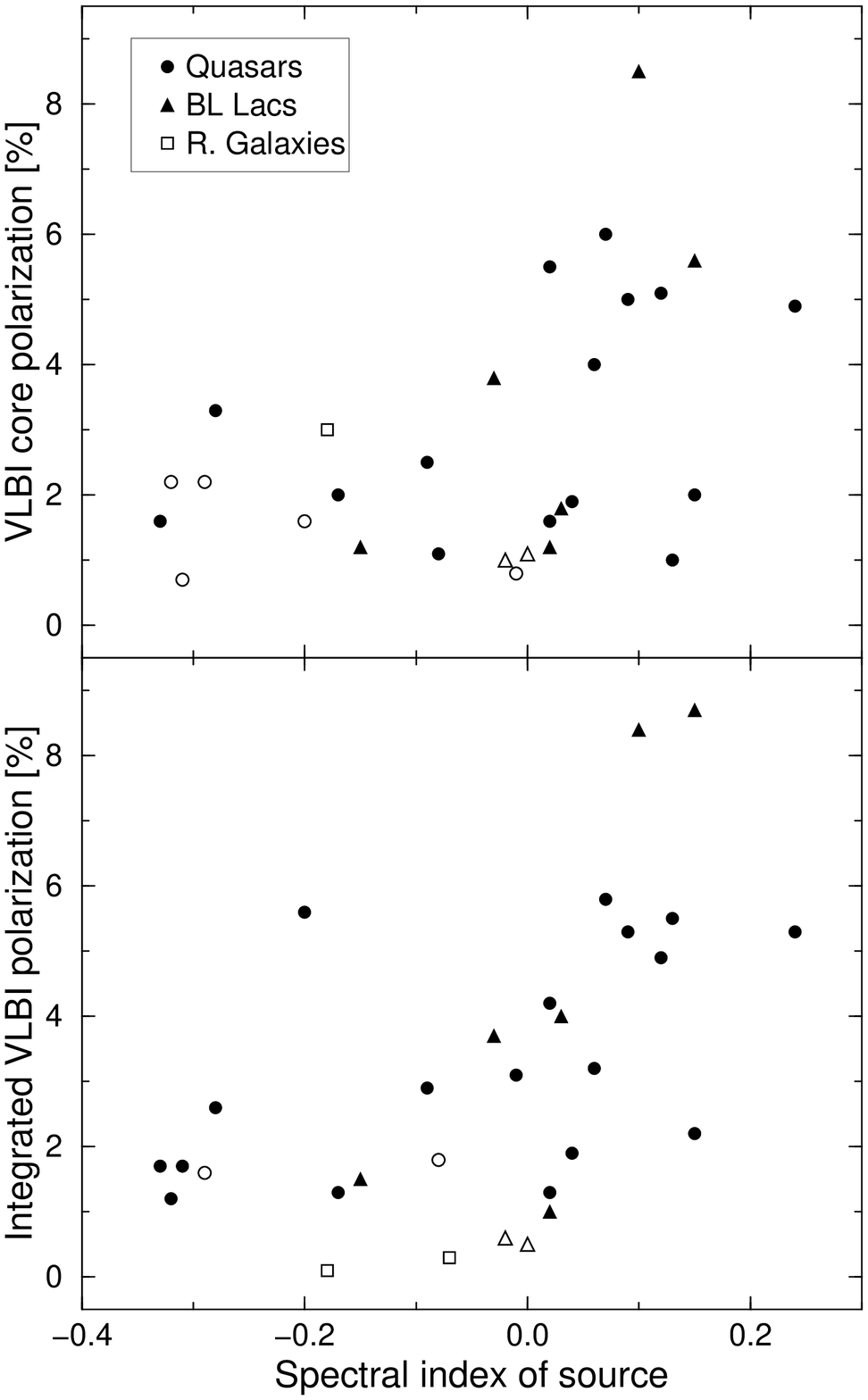}
\caption{\label{alpha__m} Top panel: fractional polarization of VLBI
core components at 43 GHz plotted against mean 5-15 GHz spectral index of the
source. Bottom panel: integrated 43 GHz VLBI fractional polarization plotted
against mean 5-15 GHz spectral index of the source. The unshaded
symbols denote upper limits in fractional polarization. }
\end{figure*}

We find that sources with flatter spectra between 5 and 15 GHz tend to
have higher core and integrated VLBI fractional polarizations at 43
GHz (Figure~\ref{alpha__m}).  The simplest means of obtaining a
flatter overall radio spectrum is to increase the dominance of the
core component, since its spectrum is usually flatter than the
jet. The correlations in Figure~\ref{alpha__m} may therefore be due to
a separate correlation between polarization and core dominance. There
is some indication that the 43 GHz core fractional polarization may be
correlated with the 5 GHz core dominance ($F_c$) in our sample, but
only at the 94.5\% confidence level, which is below our significance
cutoff.  It is not clear whether such a trend, if it exists, could be
the result of relativistic beaming effects or an intrinsic property of
the cores.

\begin{figure*}
\plotone{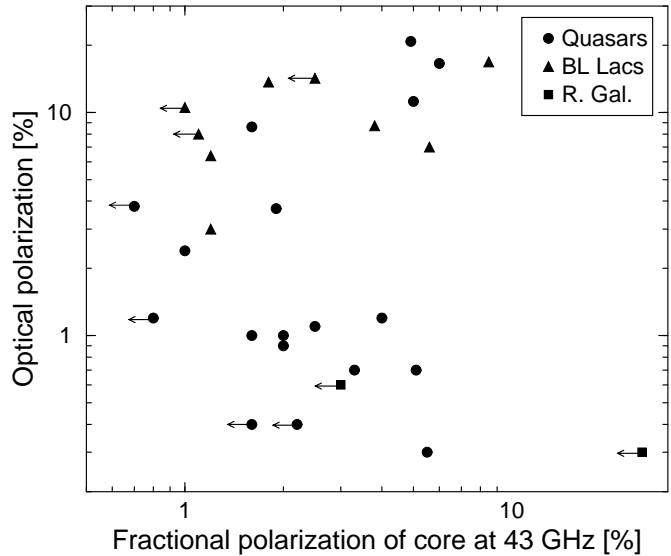}
\caption{\label{m_opt__m_core} Fractional polarization of VLBI core at
43 GHz versus optical polarization level of the source. Arrows
pointing left indicate upper limits on the core polarization.}
\end{figure*}

We find the fractional polarization of the core component at 43 GHz is also
related to the overall optical polarization of the source. 
\cite{LS00} found such a trend in a sample of low- and
high-optically polarized quasars which suggested a common (possibly
co-spatial) origin for the emission at these frequencies. In
Figure~\ref{m_opt__m_core} we plot these two quantities for the
FS-PR. The optical polarization data are tabulated in
Table~\ref{genprops}, and are taken from \cite{ILT91} with the
exception of 1954+513, whose data are taken from
\cite{WWB92}.  These authors attempted to reduce possible biases by
taking the first reliable published optical polarization measurement
for each source. We note that the optical polarization levels of
flat-spectrum radio-loud AGNs are highly variable, which will
introduce a large degree of scatter in the $y$-axis of Figure~\ref{m_opt__m_core} due
to the non-simultaneity of the radio and optical
measurements. Nevertheless, there appears to be a positive trend
present among objects with $m_{opt} \gtrsim 1\%$. There is no trend
among the low-polarization objects however. In order for them to
follow the main trend, their current optical polarization levels would
have to be significantly higher than those measured by
\cite{ILT91}.  This appears to be the case for at least one FS-PR
source: 1633+382, whose optical polarization had never exceeded 3\%
until recent observations showed variable $m_{opt}$ levels
between $\sim 1.5\%$ and 7\% (P. S. Smith 2001, private communication).

There has been a historical trend among optical astronomers to divide
quasars into low- and high-optical polarization classes at $m_{opt}
\simeq 3\%$, since most optically-selected quasars rarely exceed this
value \citep{BSW90}. This division is somewhat more problematic for
flat-spectrum, radio-loud AGNs, since their optical polarizations are
known to be highly variable. Following the standard convention, we
classify any source having a single measurement of $m_{opt} > 3\%$ as
a ``high-optical polarization quasar'' (HPQ). In order to alleviate
possible bias, we have only classified sources as ``low-polarization
radio quasars'' (LPRQs) in Table~\ref{genprops} if they have had least
three published optical polarization measurements, and have never had
$m_{opt} > 3\%$.  Unfortunately, this leaves only five genuine LPRQs
in the FS-PR, which is too small a sample for statistical
comparisons. This underscores the need for more frequent optical
polarization monitoring of larger samples of core-dominated AGNs to
better understand the connection between the radio and optical
regimes.

\subsection{Electric vector orientation of core components}
\begin{figure*}
\plotone{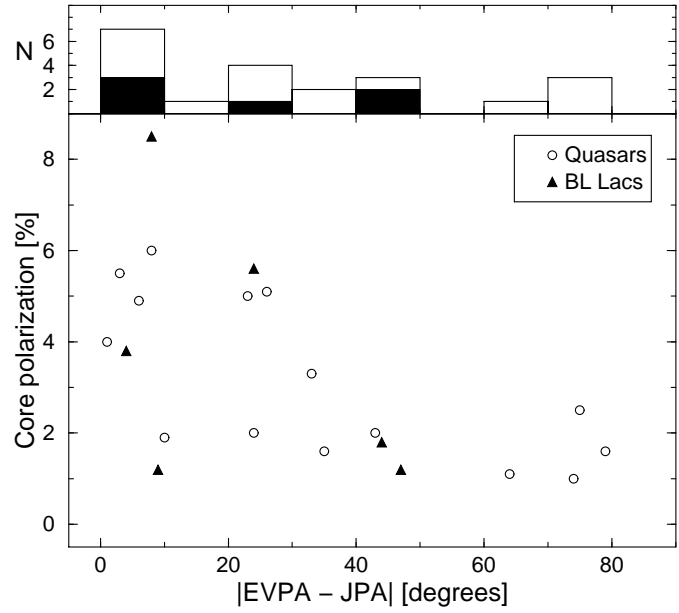}
\caption{\label{dPAcore__m_core} Fractional polarization of VLBI core
component at 43 GHz plotted against core electric vector position
angle offset with respect to the innermost jet direction. The inferred
magnetic fields of cores located towards the left hand side of the
plot are oriented perpendicular to the jet axis. }
\end{figure*}

The electric vector orientation of a polarized jet component with
respect to the local jet direction ($|{ EVPA} - { JPA}|$) is an important
quantity for testing the predictions of shock models for AGNs (e.g.,
\citealt{L80,HAA85}). These models suggest that transverse planar
shocks form in the flow and preferentially compress the perpendicular
component of the magnetic field. At high frequencies where the
emission is optically thin, this generally renders the observed electric vector
parallel to the jet axis and raises the fractional
polarization due to increased ordering of the magnetic field.  We find
evidence for such a scenario in the core components of our
sample. Figure~\ref{dPAcore__m_core} shows that as the core
polarization increases, the electric vectors tend to align with the
jet as predicted. This trend was also found in the polarized {\it jet}
components of another AGN sample at 43 GHz by \cite{LS00}. To our
knowledge this marks the first such detection in the core components
of AGNs.

The distribution of $|{ EVPA} - { JPA}|$ (top panel of
Figure~\ref{dPAcore__m_core} is somewhat skewed towards low values,
and has a 5\% probability of being drawn from a uniform
distribution, according to a K-S test. If strong transverse shocks
were dominating the polarization of all the cores, we would expect the
observed electric vectors to all lie parallel to the jet, with very
few $|{ EVPA} - { JPA}|$ misalignments greater than ten degrees. Our
measured misalignments range up to $\sim 80\arcdeg$ however.  We do
not expect the core EVPAs to be significantly affected by Faraday
rotation at 43 GHz, since extremely high RMs of $\gtrsim 25000 \; \rm
rad \; m^{-2}$ would be required to produce a significant
effect. Such values are much higher than than typical values seen in
the nuclear regions of quasars ($\sim 2000 \; \rm rad \; m^{-2}$;
\citealt*{T00}). It is possible that the limited resolution of our
images may have introduced errors in the measured $|{ EVPA} - {
JPA}|$'s due to small-scale jet bending near the core, or the blending
of unresolved components at the base of the jet. Alternatively, it may
be the case that the inner jets of some sources contain only weak
shocks that lie at oblique angles to the flow. Future high-frequency
space-VLBI observations with higher resolution should enable us to
distinguish between these possibilities.


\subsection{\label{jetdiscuss} Jet polarization properties}
The modest dynamic range of our snapshot maps does not permit
an overly detailed analysis of the faint steep-spectrum jet emission
from our sample objects. In particular, the lowest fractional
polarization we detected in any jet was approximately $2\%$.  The
properties of the jet components in our sample are tabulated
in Table~\ref{jetprops}.

\begin{figure*}
\plotone{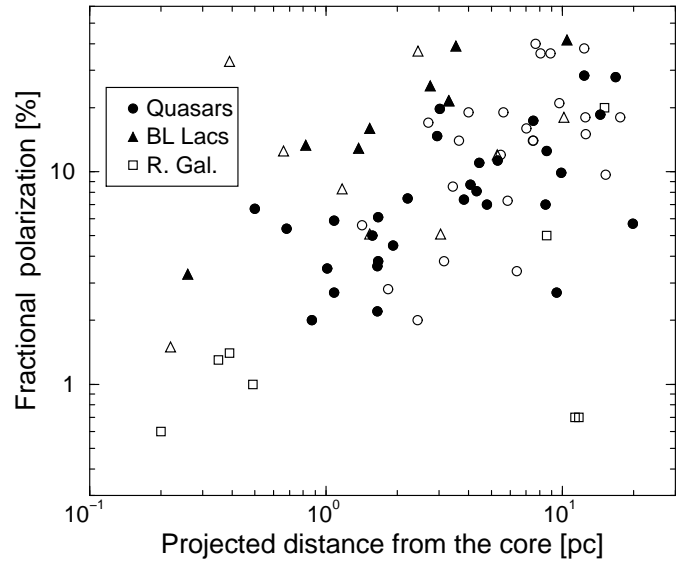}
\caption{\label{d__m_cpt} Log-log plot of fractional polarization
versus projected distance from the core for jet components in the FS-PR
sample. The filled symbols indicate components with detected
polarization, the unfilled symbols represent upper limits only.}
\end{figure*}

Despite these limitations, we have confirmed an important trend of
increasing fractional polarization with distance down the jet found by
\cite{CWRG93} (hereafter CWRG) for the PR sample at 5 GHz, and by
\cite{LS00} for a different AGN sample at 43 GHz. In
Figure~\ref{d__m_cpt} we plot fractional polarization versus distance
for the jet components (excluding the cores) in our sample. Those
components with polarized flux density greater than 5 times the rms
noise level of the $P$ image are represented by the filled symbols in
Figure~\ref{d__m_cpt}, while the unfilled symbols represent upper
limits. We will discuss the rising trend in this figure further 
in \S~5.

\begin{figure*}
\plotone{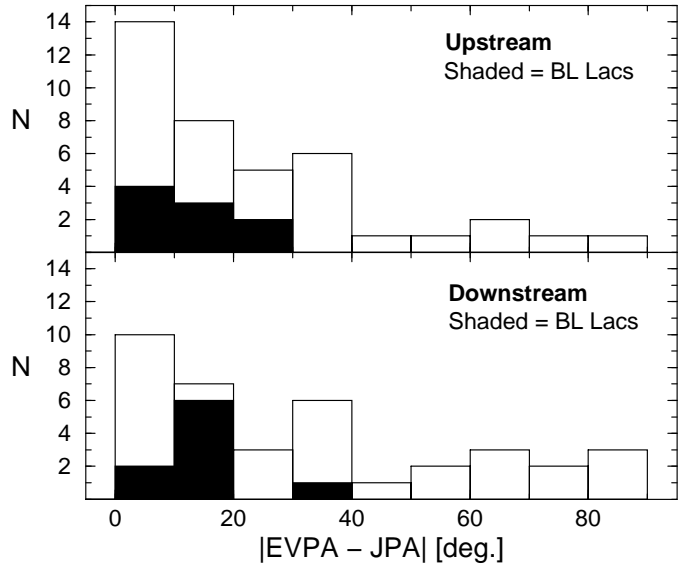}
\caption{\label{dpa_cpts} Top panel: Distribution of electric
polarization vector (EVPA) offset with respect to the local upstream
jet direction for polarized jet components (excluding cores) in the
FS-PR sample, with BL Lac components shaded. Bottom panel: EVPA offset
distribution using downstream position angles.}
\end{figure*}

We have also compared the EVPA values of the jet components to the
local jet direction for our sample. Since many of our sources undergo
bends at the position of polarized components, we define the local
direction in two ways: ${ JPA}_{in}$ as the position angle of the jet
ridge line upstream of the component, and ${ JPA}_{out}$ as the jet
position angle downstream. In those cases where the ridge line was
impossible to trace in the image due to low signal-to-noise, we used
the angle to the nearest up- or downstream component. In the upper
panel of Figure~\ref{dpa_cpts} we plot the distribution of $|{ EVPA} -
{ JPA}_{in}|$ for the polarized jet components (excluding the cores)
in the FS-PR. The distribution of $|{ EVPA} - { JPA}_{out}|$ is shown
in the lower panel. Both distributions are peaked at zero, with the
majority of components having EVPAs aligned to within $40 \arcdeg$ of
the local jet direction.

This result contrasts dramatically with that of \cite{LMG98}, who
found a uniform distribution of $|{ EVPA} - { JPA}|$ for a collection of
quasars and BL Lacs that had been observed at 22 GHz and higher at
that time. Their sample did not have well-defined selection criteria,
however, and may have suffered from unknown observational biases. 

A family of purely {\it transverse} (i.e., non-oblique) shocks cannot
reproduce our data, since their electric vectors will always be
parallel to the jet, irrespective of beaming and orientation effects
\citep{CC90}.  CWRG have suggested that quasar jets contain a strong
underlying longitudinal component to their magnetic field that
increases with distance down the jet. At the sites of shocks, the
longitudinal and transverse fields are superimposed on one another,
causing a decrease in the observed polarization. At some distance down
the jet, the underlying longitudinal field becomes strong enough to
dominate the observed polarization and flip the electric vectors
perpendicular to the jet axis. In order to account for the
intermediate values of $|{ EVPA} - { JPA}|$ in Figure~\ref{dpa_cpts}
that lie between these two extremes, some other mechanism is required.

\cite{LMG98} examined the predicted distribution of  $|{ EVPA} -
{ JPA}|$ for two possible scenarios: one in which the polarized emission
originates in moving, oblique shocks, and the other in which the
polarized components are simply enhanced jet regions with fixed
magnetic field configurations that are drawn from a random orientation
distribution. In the case of oblique shocks, they found that the
distribution of $|{ EVPA} - { JPA}|$ should be highly peaked at zero, and
will taper off sharply at high $|{ EVPA} - { JPA}|$ values. The non-oblique
model also predicts a peak near zero, but its distribution is expected
to be flatter and more uniform over the range $5\arcdeg \lesssim |{ EVPA} - { JPA}|
< 90 \arcdeg$. 

The distribution of $|{ EVPA}_{in} - { JPA}|$ in the upper panel in
Figure~\ref{dpa_cpts} is similar in shape to the oblique shock model
prediction of \cite{LMG98} for jets with upstream Lorentz factors
equal to 5. There still remain too many uncertainties in
the parameters of their model to permit detailed fitting to
our data, however.  These include the intrinsic distributions of shock
obliqueness, shock speed, viewing angle, and upstream jet speed, upon
which there are currently few observational constraints. 

There are also uncertainties regarding the true jet position angles
near the locations of polarized jet components. For example, if we
instead compare the component EVPAs to the downstream jet position
angles (lower panel of Fig.~\ref{dpa_cpts}), we cannot completely rule
out the non-oblique scenario of \cite{LMG98}. Higher-resolution
polarization images will be helpful in addressing this issue.

\subsection{Jet bending}
The combination of previous space-VLBI images of the FS-PR sample at 5
GHz \citep{LTP01} with our high-resolution 43 GHz images has enabled
us to trace the parsec-scale morphologies of these sources with
unprecedented detail. The majority of jets in the FS-PR appear
distorted by projection effects since they are likely viewed nearly
end-on. Any slight bends in the jet axis can be greatly amplified,
causing apparent bends of $90\arcdeg$ or more in some cases. For
example, the the jet of 1739+522 displays a rotation of nearly
$270\arcdeg$ in its apparent flow direction in the space of only a few
milliarcseconds (see Figure~\ref{map23} and additional images in
\citealt{L-PRI}).  

We use two measures to quantify the degree of bending in our sample,
which we tabulate in Table~\ref{genprops}. We measure both the number of
significant bends and their angular sum out to $\sim 100 \; h^{-1}$ pc
from the core using a combination of our 43 GHz images, the 5 GHz
ground and space-VLBI images from \cite{L-PRI}, and other 
VLBI images in the literature. We define a significant bend as any
change in direction of the jet greater than 
$\sim 10\arcdeg$. In order to reduce possible cosmological bias, prior to
our measurements we convolved the lower redshift sources with a larger
beam in order to simulate the effects of poorer resolution at the
median redshift of the sample ($z \simeq 1$). We did not make any
corrections for $(1+z)^4$ surface brightness dimming in the VLBI
images.  This may cause the number of bends to be overestimated in low
redshift sources, since more of their jets should be visible at a
given image sensitivity level. We find no significant difference in
the bending quantities of low ($z<1$) and high ($z>1$) redshift
sources, however. We also find no difference in the bending properties
of quasars versus BL Lac objects, even though the latter are located
at significantly closer cosmological distances.  We attribute this to
the fact that most bends tend to occur in the brightest part of the
jet near the core where the signal-to-noise ratio in the VLBI images
is high.

\subsubsection{Correlations with apparent speed}
\begin{figure*}
\plotone{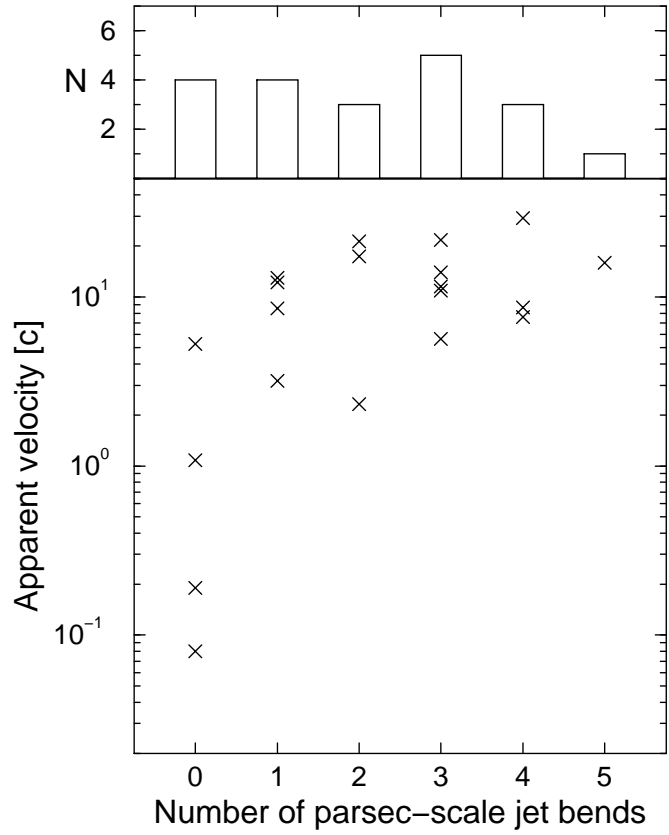}
\caption{\label{num_bend__beta_app} Plot of maximum measured apparent
jet velocity (in units of $c$) versus number of significant
jet bends ($>10\arcdeg$) out to $100\; h^{-1}$ pc from the core.}
\end{figure*}

We have compared the bending statistics of the jets in our sample to
their maximum apparent speeds as tabulated in \cite{LTP01}, and find
that in general, the straighter jets in the FS-PR tend to have slower
jet speeds. This is apparent in Figure~\ref{num_bend__beta_app}, where
we plot the number of significant bends against $\beta_{app}$. Our
Kendall's tau test shows that these quantities are correlated at the
99.39\% confidence level. A similar test of $\beta_{app}$ versus the
angular sum of the jet bends gives $97.04\%$ confidence for a
correlation. We have repeated these tests adding recent unpublished
apparent velocity data on several FS-PR objects that are part of the
VLBA 2 cm survey (K. I. Kellermann, private communication), and find
that the confidence levels increase to $99.98\%$ and $99.81\%$,
respectively.

It is not clear whether these trends may merely be due to orientation
effects that generally cause jets seen more nearly end-on to have
larger apparent bends and faster superluminal speeds. \cite{LTP01} did
not find apparent jet speed to be well-correlated with jet orientation
for the FS-PR sample, however. They attributed this to a) the narrow
predicted range of viewing angle for the sample, and b) the likelihood
that some sources are viewed inside the critical angle at which
apparent velocity is maximized. It is not obvious what intrinsic
processes could lead to an intrinsic connection between jet speed and
bending. Current theory suggests that jet bends can arise from
Kelvin-Helmholtz instabilities that are driven by an external periodic
process \citep{H84}. Although the evolution of these instabilities is
highly dependent on the jet Mach number, it is not well-known how the
latter quantity is related to the observed VLBI component speeds.
More theoretical work is needed to determine how jet speed influences
the observed polarization of shocked relativistic jets with intrinsic
bends. Detailed three-dimensional magnetohydrodynamical computer
simulations should be extremely useful in this regard.

\subsubsection{\label{realign} Bending morphology}
A large fraction of the jets in the FS-PR sample exhibit a peculiar
morphology in which the jet starts out with a particular flow
direction, undergoes a bend, and then resumes with nearly exactly the
original flow direction. Approximately half of the sources (16 of 36)
we observed at 43 GHz display this ``bend and re-align'' (BAR)
morphology, which we indicate in column 10 of Table~\ref{genprops}. A
good example in our survey can be found in the jet of 1828+487
(Figure~\ref{map28}), where the position angle between components D
and E is very similar to that between A and B, despite a large bend in
between these regions. Our two-sample tests do not reveal any strong
differences in the overall properties of the BAR versus non-BAR
sources in our sample. 

\begin{figure*}
\plotone{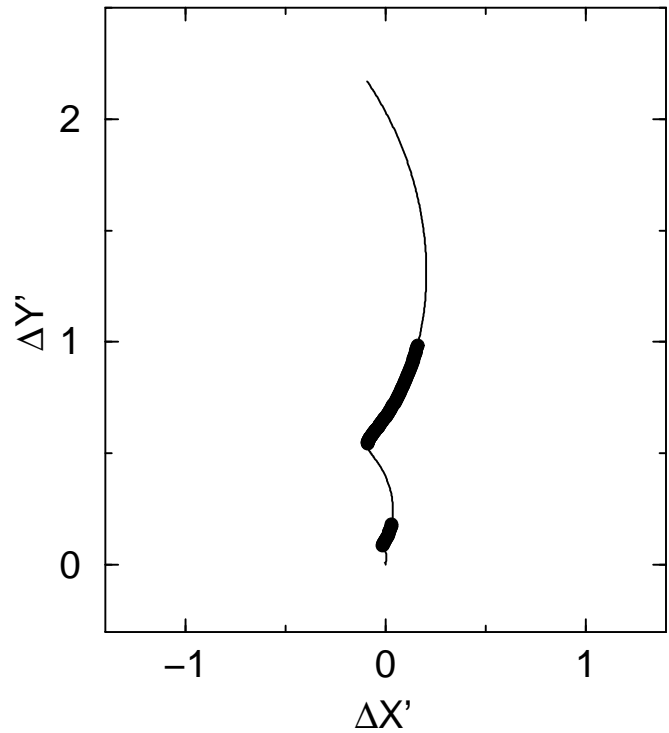}
\caption{\label{helix} Projection of a damped helical jet onto the sky
plane for an observer located at a viewing angle of 6 degrees from the
helix axis. The regions marked in bold subtend a viewing angle of less
than 5.5 degrees to the observer, and will be significantly Doppler
boosted if the bulk flow follows the streamlines and is
relativistic. The position angle of the jet is similar in both boosted
regions, which is reminiscent of the morphology seen in many of the
jets in the FS-PR sample.}
\end{figure*}

The BAR morphology is highly reminiscent of a jet in which the
streamlines follow a low-pitch helical pattern, once relativistic
beaming effects have been taken into account.  \cite{GAM94} have
demonstrated that the emission is maximized in regions of the jet that
have small angles to the line of sight. We illustrate this in
Figure~\ref{helix}, where the thin line shows the projection of a
damped low-pitch helix onto the sky plane (denoted by primes). The
axis of the helix lies along the $\hat z$ direction in a right-handed
coordinate system. The observer is located at a viewing angle of
$6\arcdeg$ from the $\hat z$ axis, and an azimuth direction of
$45\arcdeg$ from the $\hat y -\hat z$ plane, measured counterclockwise
looking down the $+\hat z$ axis toward the origin. The helix itself
has a half-opening angle of $1\arcdeg$, and an increasing wavelength
of the form $\lambda = \lambda_o z^{\epsilon}$, where we have chosen
$\lambda_o = 0.2$ and $\epsilon = 0.3$ in arbitrary units. The regions
marked in bold have smaller viewing angles than the rest of the jet,
and thus will appear much brighter and more beamed. In VLBI images of
limited sensitivity, it is possible that these may be the only jet
regions that are detected.

We have chosen this particular model merely to illustrate how beaming
and orientation effects can conspire to produce BAR morphologies in
AGN jets. We note that our model assumes that the bulk motions of the
jet plasma follow helical streamlines, and are not merely
ballistic. Evidence for this can be found in several sources
(0859+470, 1641+399, 1642+390, 1803+784, and 2200+420) in which the
polarization vectors tend to follow changes in the jet ridge line
direction.

\section{Comparison of BL Lac object and quasar polarization properties}

A major unresolved issue in AGN research is the relation of
high-optical polarization quasars to their weak-lined cousins, the BL
Lacertae objects. The separation of these two classes is still a
matter of some debate, as the original classification criterion for BL
Lacs (strongest rest frame emission line width narrower than 5 \AA; e.g., 
\citealt*{SFK91}) does not appear to reflect any natural division in the
observed line width distribution of blazars \citep{SF97}. A further
complication arises in that the emission line widths of many blazars
(including BL Lac itself; \citealt*{VOT95}) are known to vary with
time. Nevertheless, several studies have shown marked differences
in the properties of broad- and narrow-lined blazars.  These include
their spectral properties in the sub-mm
\citep{GSH94} and x-ray \citep{Pad92} bands, and their jet
polarization characteristics on parsec scales (CWRG).

CWRG obtained 5 GHz VLBI polarization data on 24 sources in the
Pearson-Readhead sample, 20 of which are members of the FS-PR.  They
found that the overall jet EVPAs of BL Lacs tended to be much more
aligned with the jet than those of quasars.  Compared to our 43 GHz
observations, their data were more sensitive to weakly polarized
emission further down the jet, but had significantly poorer spatial
resolution and were more influenced by Faraday rotation and
de-polarization effects. Our data are more sensitive to polarization
structure closer to the base of the jet. In this section we compare
the 43 GHz VLBI polarization properties of the BL Lacs and quasars in
the FS-PR sample to those obtained by CWRG at 5 GHz. Our BL Lac
classifications are based on the catalog of \cite{PG95}, and are
identical to those of CWRG.
 
\subsection{Core components}
CWRG found that at 5 GHz, the VLBI core components of quasars tended to
have smaller fractional polarizations than those of BL Lacs. The
$|{ EVPA}_{core}-{ JPA}|$ distributions of both classes were randomly
distributed, indicating no preferred orientation of the core magnetic
field with respect to the jet.  We find a similar result at 43 GHz
regarding the core EVPAs (see Figure~\ref{dPAcore__m_core}), but
find no significant differences in the core polarization levels of the
two classes (top panel of Figure~\ref{m_plots}). It is possible that
CWRG observed a blend of polarization from components located near the
core. This may have reduced the observed polarization of the quasar cores,
which are generally located at larger cosmological distances than the
BL Lacs.  Higher resolution mm- or space-VLBI observations are
required to determine whether this is also the case for our 43 GHz
observations.

\subsection{\label{BL_qso_jet}Jet components}
CWRG also found significant differences in the magnetic field
properties of BL Lacs and quasar jets. They showed that BL Lacs
tend to have electric polarization vectors that line up parallel
to the jet, while quasars have predominantly perpendicular EVPA
orientations.  We note that CWRG used a vector-averaged EVPA for the
jet and compared it to the position angle of the component with the
largest flux density. This method is not appropriate for our data,
due to the large number of small scale jet bends present in many
sources. We have instead plotted the distribution for all polarized
components, compared to the local upstream and downstream jet
directions (Figure~\ref{dpa_cpts}, upper and lower panels respectively).

There do appear to be differences in the $|{ EVPA} - { JPA}|$ distributions
for BL Lacs and quasars at 43 GHz, but not as strong a level as
reported by CWRG. A Kolmogorov-Smirnov (K-S) test on the BL Lac and
quasar distributions in the lower panel of Figure~\ref{dpa_cpts} shows
that they differ at the 98.88\% confidence level, while in the upper
panel, the confidence level is only 83.53\%. These lower confidence
levels are a result of a much stronger tendency for quasar jets to
have electric vectors aligned with the jet at 43 GHz than at lower
frequencies. 

We confirm CWRG's finding that BL Lac jet EVPAs are preferentially
aligned with the local jet direction. There are no BL Lac jet
components with $|{ EVPA} - { JPA}| > 40\arcdeg$, irrespective of
whether we use the upstream or downstream JPAs. If the intrinsic
distribution of $|{ EVPA} - { JPA}|$ were in fact uniform between 0
and $90\arcdeg$, the binomial probability of finding all nine BL Lac
components in the first four bins of Figure~\ref{dpa_cpts} by chance
is only $(4/9)^9 = 0.068\%$. 

\begin{figure*}
\plotone{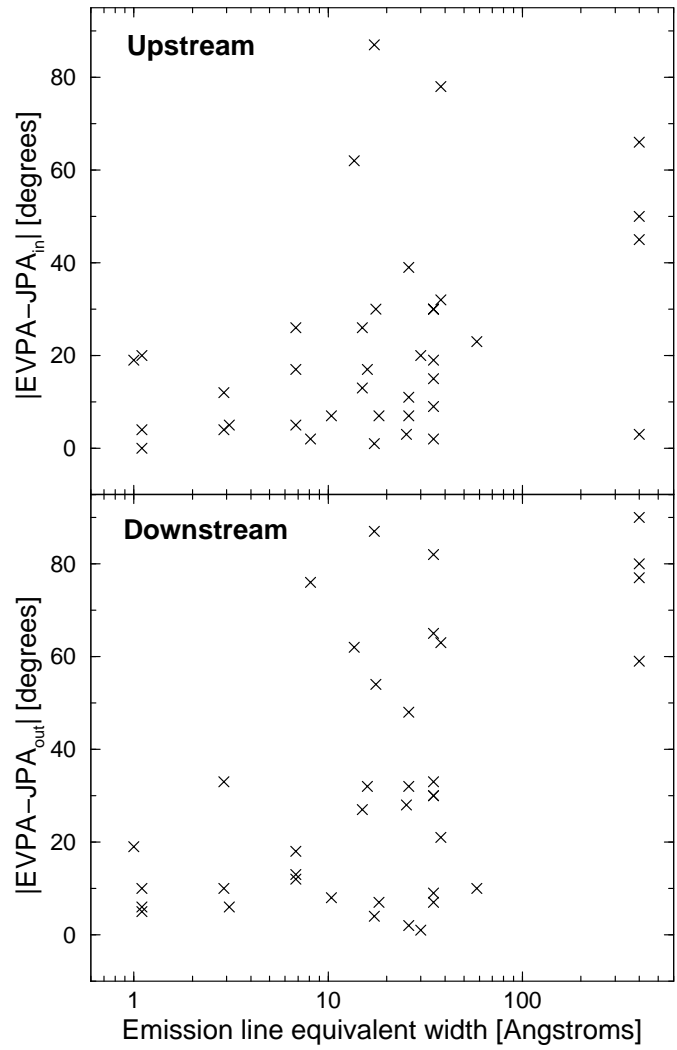}
\caption{\label{ew__dpa} Top panel: EVPA offset of polarized jet
components with respect to the upstream jet direction, plotted against
rest frame equivalent width of the broadest permitted line in the
source. Bottom panel: same plot using downstream jet directions.}
\end{figure*}

The connection between jet magnetic field properties and optical line
emission is also apparent in Figure~\ref{ew__dpa}, in which we plot
the EVPA offset of each jet component against the rest frame
equivalent width of the broadest permitted line in the source's
optical spectrum. The latter data were tabulated by \cite{LZR96} and
\cite{LTP01}. It is evident that all of the
components with $|{ EVPA} - { JPA}|> 40\arcdeg$ are found in sources
having emission lines broader than $\sim 8$ \AA.

\begin{figure*}
\plotone{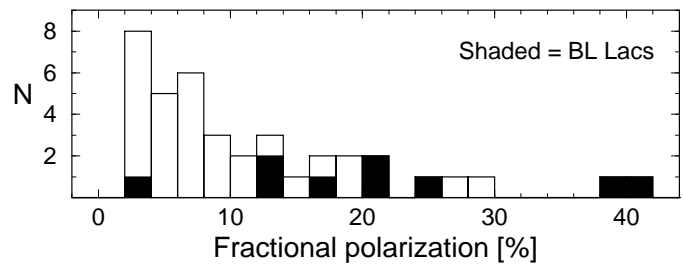}
\caption{\label{cpt_m} Distribution of 43 GHz fractional polarization for jet
components with detected polarized emission, with those of BL Lacs
shaded.}
\end{figure*}

Despite finding apparent differences in electric field orientation of BL
Lac and quasar jets, CWRG did not find any significant differences in
their degree of polarization. A K-S test on our 43 GHz data
(Figure~\ref{cpt_m}) gives a 97.04\% probability that the $m_{cpt}$
distributions of the two classes are different, which is slightly
below our cut-off level of $98\%$ at which we call a correlation
significant.  Evidence of a difference in 43 GHz fractional
polarization levels can, however, be found in the plot of $m_{cpt}$
versus projected distance from the core (Figure~\ref{d__m_cpt}). This
plot shows that {\it at a given projected distance from the core}, the
fractional polarizations of BL Lac components are generally higher
than those of quasars.

\subsection{\label{longitud} Longitudinal magnetic field model}
CWRG interpreted their polarization findings at 5 GHz in terms of a
model that incorporates a longitudinal magnetic field component
present in the jets of quasars, but not in those of BL Lacs. They
claimed that the underlying jet polarization in quasars is very strong
and increases down the jet due to shear with the external medium. The
observational signatures of this model are a general tendency in
quasars for the fractional polarization to increase and the EVPAs to
become more perpendicular to the jet axis as you move downstream from
the core.

Although the longitudinal field model does appear valid at large
distances downstream, several difficulties arise when we attempt to
apply it to regions closer to the core. First, we find that at 43 GHz,
quasars in fact contain many polarized components with electric
vectors {\it parallel} to the jet, suggesting that shocks dominate any
underlying longitudinal field in these regions.  Second, we find no
tendency for the EVPAs to become more perpendicular to the jet
with increasing distance from the core. Indeed, the four quasar
components with perpendicular EVPA alignments ($> 60\arcdeg$ from
the upstream jet direction) are all located within 2 parsecs
(projected) of the core, and not far downstream. It would appear
likely then that in the jet regions within several milliarcseconds of
the core, the longitudinal shear component of the magnetic field has
not yet fully developed. In this case the trend of increasing
polarization may instead result from some other mechanism, such as an
increase in shock strength or viewing angle with core
distance. \cite{LS00} point out that a jet which is bending away from
the observer can create an increasing trend of fractional polarization
with distance, given the right geometry. Since apparent velocities are
also dependent on viewing angle, it should be possible to distinguish
between an increase in shock strength versus an increase in viewing
angle by monitoring the superluminal speeds and polarization
properties of components as they move down the jet.

In summary, we find evidence for fundamental differences in the inner
jet properties of BL Lac and quasars that need to be investigated in
more detail using higher-sensitivity images taken at multiple
epochs. Such data can yield good constraints on the strengths and
viewing angles of relativistic shocks (e.g., \citealt*{WCRB94}), and
would clarify considerably our observational picture of the magnetic
field structures in AGN jets.

\section{Conclusions}
We have carried out the first 43 GHz VLBI polarization survey of a
complete sample of active galactic nuclei at a resolution of $\sim
0.2$ milliarcseconds, making this the highest resolution imaging
survey of a complete AGN sample made to date. Our sample consists of a
flat-spectrum subset of the well-known Pearson-Readhead survey
(FS-PR), which has been thoroughly studied at a variety of wavelengths
and resolution levels. We have investigated the jet magnetic field
properties of these objects on parsec-scales close to the central
engine.  The main results of our study are as follows:

1. The morphologies of bright, flat-spectrum AGNs at 43 GHz generally
consist of a bright core component located at the base of a faint
one-sided jet. On average, the core component accounts for $\sim 60\%$
of the total VLBI flux at this frequency. Its luminosity is
well-correlated with numerous other luminosity parameters, including
the total source luminosity in soft x-rays, in the optical, and at 5
GHz.

2. The ratio $R$ of VLBI core-to-remaining VLBI flux is 
well-correlated with several statistical indicators of relativistic
beaming, namely optical polarization, 5 GHz radio variability
amplitude, and variability brightness temperature.

4. The linear polarizations of the core components in the FS-PR sample
range up to $\sim 9\%$, with typical values of $\sim 1.5\%$. Although
the jets generally show higher fractional polarization levels than the
cores, the bulk of the polarized emission on parsec-scales comes from
the core component. In contrast to previous 5 GHz VLBI measurements,
we find the cores of BL Lacertae objects and quasars to have similar
polarization levels at 43 GHz. Also, sources with more highly
polarized cores tend to have flatter spectral indices measured between
5 and 15 GHz.

5. There is a tendency for the more highly polarized cores to have electric
field vectors that lie preferentially along the jet axis. This
behavior is consistent with  shocks in the inner jet flow
that preferentially compress the perpendicular component of the
magnetic field.

6. Among the sources with optical polarizations ($m_{opt}$) exceeding
$\sim 1\%$, there is a positive trend of $m_{opt}$ with 43 GHz core
polarization, which suggests a common origin for the emission at these
two frequencies. We find no such trend among the lower-optically polarized
objects.

7. The fractional polarization of jet components tends to increase
with projected distance down the jet in a similar manner for both BL
Lacs and quasars. However, at a given distance from the core, the jet
components of BL Lacs are more polarized, and have magnetic fields
preferentially aligned perpendicular to the jet.

8. The vast majority of polarized components located in the jets of
both BL Lac objects and quasars have electric field vectors that lie
within $40\arcdeg$ of the local jet direction. The distribution of
electric vector offsets with respect to the jets in the entire sample
is consistent with theoretical predictions for an ensemble of moving
oblique shocks.

9. We find that straighter jets tend to have smaller apparent jet
speeds than sources that display significant jet bending. Nearly half of
the jets in the FS-PR sample display a ``bend-and-realign'' (BAR)
morphology that is a predicted feature of jet plasma streaming
relativistically along a helical path. We find support for such
streaming motions in several objects whose electric vectors closely
follow significant changes in the jet direction.

\acknowledgments
 The author thanks Bob Preston, Glenn Piner, and Steven Tingay for
 helpful discussions during the planning stages of this project, and
 to Alan Marscher and an anonymous referee for comments on the
 manuscript. Thanks also go to Harri Ter\"asranta for providing
 flux-density data, and Tim Cawthorne, Alan Marscher, Svetlana
 Jorstad, and Ken Kellermann for providing VLBI data in advance of
 publication.

 This research was performed in part at the Jet Propulsion Laboratory,
 California Institute of Technology, under contract to NASA, and has
 made use of data from the following sources:

 The NASA/IPAC Extragalactic Database (NED), which is operated by the
 Jet Propulsion Laboratory, California Institute of Technology, under
 contract with the National Aeronautics and Space Administration.

 The University of Michigan Radio Astronomy Observatory, which
 is supported by the National Science Foundation and by funds from the
 University of Michigan.

\clearpage
\begin{deluxetable}{lllccrrrcrc}
\tablecolumns{11}
\tablecaption{\label{genprops}General properties of Pearson-Readhead AGNs}
\tablewidth{0pt}
\tablehead{ \colhead{IAU} &\colhead{Other}  &\colhead{Opt.}& & \colhead{log}
&  &  & &  & & \colhead{ Bend \&} \\
\colhead{Name} &\colhead{Name}& \colhead{Class} & \colhead{z}
& \colhead{${L_{tot}}$} &\colhead{$m_{tot}$}    &  \colhead{$m_{opt}$}
& \colhead{$JPA$} & \colhead{$N_{bend}$}  
& \colhead{$\Sigma_{bend}$}  & \colhead{Revert} \\
\colhead{ [1]} & \colhead{ [2]} & \colhead{ [3]}  &
\colhead{ [4]} & \colhead{ [5]} & \colhead{ [6]} &
\colhead{ [7]}& \colhead{ [8]}  & \colhead{ [9]} &\colhead{ [10]}
&\colhead{ [11]}}

\startdata
\sidehead{Flat-spectrum (FS-PR) sample objects}
0016+731&   \n      &Q&1.781 &27.78& 2.2&  0.9&   132&  2  & 66 &N  \\
0133+476&OC 457     &HPQ&0.859 &27.95& 5.3& 20.8&   330& 0 &0&N\\
0212+735&   \n      &HPQ&2.367 &28.56& 4.9&  0.7&   121&  4  & 159  &Y\\
0316+413&3C 84      &RG &0.017 &24.52&$<0.1$&  0.6&   176&0&0 &N  \\
0454+844&   \n      &BL &0.112 &24.54&$<2.4$& 14.2&   177&0&0&Y  \\
0723+679&3C 179     &LPRQ&0.844 &27.12& 1.3&  1.0&   256&  1  & 24 &N \\
0804+499&OJ 508     &HPQ&1.432 &27.59& 1.3&  8.6&   127&  1 & 11 & N \\
0814+425&OJ 425     &BL &0.245 &26.13& 3.7&  8.7&   103&  2 & 142 &  Y \\
0836+710&4C 71.07   &Q&2.180 &28.38& 1.7&  1.0&   201&    3   & 45 & Y  \\
0850+581&4C 58.17   &LPRQ&1.322 &27.21&$<1.6$&  0.4&   227& 4& 147&Y \\
0859+470&4C 47.29   &Q&1.462 &27.46& 2.6&  0.7&   357&  2&  43 & N  \\
0906+430&3C 216     &HPQ&0.670 &26.81& 1.7&  3.8&   151&0 &0& Y  \\
0923+392&4C 39.25   &LPRQ&0.699 &28.11& 5.6&  0.4&    93&2  & 59 & Y  \\
0945+408&4C 40.24   &LPRQ&1.252 &27.86& 4.2&  0.3&   137&4  & 65 &Y \\
0954+556&4C 55.17   &HPQ&0.900 &   \n&  \n&  6.4&   191& 1&106& N  \\
0954+658&   \n      &BL &0.367 &26.19& 1.0&  6.4&   311& 1 & 41 & N  \\
1624+416&4C 41.32   &Q&2.550 &28.00& 1.2&   \n&   262&  2  & 193 & N \\
1633+382&4C 38.41   &HPQ&1.807 &28.40& 2.9&  1.1&   279&  5  & 194 & N\\
1637+574&OS 562     &Q&0.749 &27.55& 5.5&  2.4&   200&0 &0& Y \\
1641+399&3C 345     &HPQ&0.595 &27.90& 5.3& 11.2&   284& 4 & 203 & Y \\
1642+690&4C 69.21   &HPQ&0.751 &27.31& 5.8& 16.6&   158& 3 & 109  & Y \\
1652+398&MK 501     &BL &0.033 &24.10& 1.5&  3.0&   160& 3  & 225  & Y \\
1739+522&4C 51.37   &HPQ&1.381 &27.65& 1.9&  3.7&   204& 2 & 185 & N \\
1749+701&   \n      &BL &0.770 &26.82& $<0.6$& 10.5&275& 3 & 248 & N \\
1803+784&   \n      &BL &0.680 &27.40& 8.7&  7.0&   291& 1& 25 & N \\
1807+698&3C 371     &BL &0.050 &24.68&$<0.5$&  8.0&   252& 1  & 10 & N\\
1823+568&4C 56.27   &BL &0.663 &27.24& 8.4& 16.8&   201& 2  & 40 & Y \\
1928+738&4C 73.18   &LPRQ&0.302 &26.83& 3.1&  1.2&   166&   3 & 164 &Y \\
1954+513&OV 591     &Q&1.223 &27.67& 3.2&  1.2&   307&  0 &0& Y \\
2021+614&OW 637     &RG &0.228 &26.21&$<0.3$&  0.3&   214& 0&0& N \\
2200+420&BL Lac     &BL &0.069 &25.81& 4.0& 13.7&   209&   1  & 65 & N \\
2351+456&4C 45.51   &Q &1.986 &28.21& 1.8& \n& 321&   1 & 19 & N \\

\sidehead{Other PR sample objects}
0153+744&   \n      &Q&2.338 &27.12&$<$2.3&  1.1&    68&    2&151&N \\
0538+498&3C 147     &LPRQ&0.545 &26.39&$<$1.9&  2.3&   237& 0&0 & N  \\
0711+356&OI 318     &Q&1.620 &27.08&$<$1.6&  1.0&   329&  0&0& N\\
1828+487&3C 380     &LPRQ&0.692 &27.27& 3.8&  0.2&   311&  4 & 143 & Y \\
\enddata 

\tablecomments{\scriptsize  All quantities are derived assuming
$h = 0.65$, $q_o = 0.1$, and zero cosmological constant.\\
Col. (1).--- IAU source name. \\
Col. (2).--- Alternate name. \\
Col. (3).--- Optical classification: RG = radio galaxy, BL = BL Lac,
HPQ = high (optical) polarization radio quasar, LPRQ = low (optical)
polarization radio quasar, Q = radio quasar with unknown optical polarization \\
Col. (4).---  Redshift from \cite{LZR96}.  \\
Col. (5).---  Total VLBI luminosity at 43 GHz [$\rm W\; Hz^{-1}$]. \\
Col. (6).---  Total VLBI integrated fractional polarization of source
at 43 GHz [\%]. \\
Col. (7).---  Total optical fractional polarization of source [\%]. \\
Col. (8).---  Innermost jet position angle on parsec scales [degrees]. \\
Col. (9).--- Number of significant jet bends (greater than $10\arcdeg$) within  $100\; h^{-1}$ projected parsecs of the core. \\
Col. (10).--- Sum of significant bend angles within  $100\; h^{-1}$ projected parsecs of the core [degrees]. \\
Col. (11).--- Jet bend and re-align (BAR) morphology (see \S~\ref{realign}). \\
}

\end{deluxetable}

\clearpage
\begin{deluxetable}{lllrrrrrrl}
\tabletypesize{\scriptsize}

\tablecolumns{10}
\tablecaption{\label{mapstats}Summary of 43 GHz image parameters}
\tablewidth{0pt}
\tablehead{
\colhead{\ Source} & \colhead{Obs. Date} & \colhead{Type} &
\colhead{\ Beam}&  \colhead{\ PA} &
\colhead{\ Flux} & \colhead{\ EV} & \colhead{\ RMS} &
\colhead{\ Peak} &
\colhead{\ Contour levels}\\
\colhead{\ [1]} & \colhead{\ [2]} & \colhead{\ [3]}  &
\colhead{\ [4]} & \colhead{\ [5]} &
\colhead{\ [6]} & \colhead{\ [7]}  & 
\colhead{\ [8]} & \colhead{\ [9]} & \colhead{\ [10]}}

\startdata  
0016+731  & 1999 Apr 6 &  IPOL & 0.26 x 0.24 & 58 & 668 & \n & 1.0 & 598 & $-0.4$, 0.4, 0.8, 1.6, 3.2 \\
    &     &   PPOL &     \n       & \n  & 15 & 88 & 0.8 &  12 & \n \\
0133+476  & 2000 Sep 10 & IPOL &  0.32 x 0.20 &  34 & 4464 &  \n & 2.1
 & 4049&  $-0.125$, 0.125, 0.25, 0.5, 1, 2\\
    &     &   PPOL &     \n       & \n  &  238 & 306 & 1.1 &    198&\n \\
0153+744  & 2000 Jan 5 &  IPOL & 0.29 x 0.26 & $-85$ & 87 & \n & 0.6 &
 81 & $-1.75$, 1.75,  3.5, 7, 14, 28,  56 \\
    &     &   PPOL &     \n       & \n  & $<2.0$ & 35  & 0.4  & 3.4 & \n \\
0212+735  & 1999 Apr 6 &  IPOL & 0.22 x 0.19 & 63 & 1226 & \n & 1.7 &
 451 & $-1$, 1, 2, 4, 8, 16, 32, 64 \\
    &     &   PPOL &     \n       & \n & 60 & 117 & 1.1 & 25 & \n \\
0316+413  & 1999 Apr 6 &  IPOL & 0.30 x 0.20 & $-7$ & 4362 & \n & 2.9
& 973 & $-1.1$, 1.1, 2.2, 4.4, 8.8, 17.6, 35.2, 70.4 \\
    &     &   PPOL &     \n       & \n & $<5.5$ & 47 & 1.1 &  7.0 & \n \\
0454+844  & 2000 Jan 5 &  IPOL & 0.34 x 0.31 & $-19$ & 106 & \n & 0.5
&  81 & $-1.75$, 1.75, 3.5, 7, 14, 28,  56 \\
    &     &   PPOL &     \n       & \n  & $<2.5$ & 35  & 0.5 & 2.7 & \n \\
0538+498  & 1999 Apr 6 &  IPOL & 0.28 x 0.22 & $-8$ & 286 & \n & 1.4 &
172 & $-2.25$, 2.25, 4.5, 9, 18, 36, 72 \\
    &     &   PPOL &     \n       & \n & $<5.5$ & 70 & 1.1 &   7.7 & \n \\
0711+356  & 2000 Jan 5 &  IPOL & 0.40 x 0.23 & $-17$ & 122 & \n & 0.6
&71 & $-2$, 2, 4, 8, 16, 32,  64 \\
    &     &   PPOL &     \n       & \n  & $<2.0$ & 18  & 0.4 &  3.5 & \n \\
0723+679 & 2000 Sep 10 & IPOL &  0.39 x 0.20 &  90 &  712 & \n  & 1.1
&   637&   $-$0.4, 0.4, 0.8, 1.6, 3.2, 6.4, 12.8, 25.6\\
    &     &   PPOL &      \n      &  \n &   9.5  & 61  & 0.9 &    13&\n \\
0804+499  & 1999 Apr 6 &  IPOL & 0.29 x 0.20 & $-29$ & 721 &\n & 0.7 &
697 & $-0.3$, 0.3, 0.6, 1.2, 2.4, 4.8, 9.6 \\
    &     &   PPOL &     \n       & \n & 9.5 & 78  & 0.8 & 11 & \n \\ 
0814+425  & 2000 Jan 5 & IPOL & 0.31 x 0.24 & 0 & 862 &  \n &0.7 &
 724 & $-0.25$, 0.25, 0.5, 1, 2, 4, 8\\
    &     &   PPOL &     \n       & \n  & 32 & 47  & 0.6 &   27 & \n \\
0836+710  & 1999 Apr 6 &  IPOL & 0.25 x 0.23 & $-53$ & 1434 & \n & 1.1
&  1091 & $-0.3$, 0.3, 0.6, 1.2, 2.4, 4.8, 9.6, 19.2, 38.4, 76.8\\ 
    &     &   PPOL &     \n       & \n & 24 & 70  & 0.7 &  17 & \n \\
0850+581 & 2000 Sep 10 & IPOL &  0.45 x 0.21 & $-$81 &  278 & \n  &
1.7 &   206&  $-$1.5, 1.5, 3, 6, 12 \\
    &     &   PPOL &    \n        &  \n &  $<4.5$& 31  & 0.9 &     5.0&\n \\
0859+470  & 2000 Jan 5 & IPOL & 0.37 x 0.26 & $-2$ & 389 & \n &0.6 &
 253 & $-0.75$, 0.75, 1.5, 3, 6, 9, 18, 36 \\
    &     &   PPOL &     \n       & \n  & 10 & 18 & 0.4 & 8.5 & \n \\
0906+430  & 1999 Apr 6 &  IPOL & 0.41 x 0.38 & $-85$ & 498 & \n & 0.9
& 375 & $-0.75$, 0.75, 1.5, 3, 6, 9, 18, 36, 72 \\
    &     &   PPOL &     \n       & \n & 8.5 & 35  & 0.5 &  10 & \n \\
0923+392\tablenotemark{a} & 1999 Jan 12 & IPOL &  0.29 x 0.17 & $ -1$   &  6986  & \n &
1.3 &   2204 & $-0.4$, 0.4, 0.8, 1.6, 3.2, 6.4, 12.8, 25.6, 51.2 \\
 &   &  PPOL      &    \n          &    \n       &  394  & 233 &  1.5 &    160 &\n \\
0945+408  & 2000 Jan 5 & IPOL & 0.24 x 0.17 & 0 & 1176  & \n &1.4 &
 670 & $-0.6$, 0.6, 1.2, 2.4, 4.8, 9.6, 19.2, 38.4 \\
    &     &   PPOL &     \n       & \n  & 49 & 140 & 1.1 &  36 & \n \\
0954+658  & 1999 Apr 6 &  IPOL & 0.25 x 0.21 & $-74$&  448 & \n & 0.9 &
 402 & $-0.7$, 0.7, 1.4, 2.8, 5.6, 11.2, 22.4 \\
    &     &   PPOL &     \n       & \n & 4.5 &  47 & 0.9 & 6.9 & \n \\
1624+416  & 1999 Apr 6 &  IPOL & 0.38 x 0.27 & $-17$& 339 &\n &  1.0 &
227 & $-0.65$,0.65,  1.3, 2.6, 5.2, 10.4, 20.8 \\
    &     &   PPOL &     \n       & \n & 4.1 & 35 & 1.0 &  7.0 & \n \\
1633+382\tablenotemark{a} & 1999 Jan 12 & IPOL &  0.33 x 0.21 & $ -13 $&  2286 &\n  &
0.7 &  1736& $-0.125$, 0.125, 0.25, 0.5, 1, 2, 4, 8, 16, 95 \\
   &     &  PPOL &    \n            &   \n           & 67 &  156 &  0.9 &    43&\n \\
1637+574  & 1999 Apr 6 &  IPOL & 0.26 x 0.22 & 4 & 1898 & \n & 0.7 &
 1019 & $-0.15$, 0.15, 0.3, 0.6, 1.2, 2.4, 4.8, 57.6 \\
    &     &   PPOL &     \n       & \n & 104 & 88  & 0.7 &  42 & \n \\
1641+399 & 1998 Jul 31 & IPOL & 0.24 x 0.17 & $-16$ & 8168 & \n & 1.8
&  4887 & $-0.15$, 0.15, 0.3, 0.6, 1.2, 2.4, 4.8, 9.6\\
 &       &  PPOL &  \n              &  \n          &   433  & 205 & 1.2 &  252 & \n \\
1642+690 & 2000 Sep 10 & IPOL &  0.25 x 0.21 & $-$30 & 1232 & \n  &
0.7 & 1061& $-$0.2, 0.2, 0.4, 0.8, 1.6, 3.2\\
    &     &   PPOL &      \n      &  \n & 72   & 77  & 0.5 &     65&\n \\
1652+398 & 1997 May 26 & IPOL & 0.35 x 0.22 & $-9$ & 461   & \n &
0.6& 231& $-0.7$, 0.7, 1.4, 2.8, 5.6  \\
  &       &  PPOL &  \n              &  \n          & 6.7 & 17 & 0.6 &  6.7 & \n \\
1739+522 & 2000 Sep 10 & IPOL &  0.30 x 0.20 & $-$19 &  782 &  \n &
0.7 &    657&  $-$0.25, 0.25, 0.5, 1, 2\\
    &     &   PPOL &      \n      &  \n &  15  & 57  & 0.5 &     16&\n \\
1749+701 & 2000 Sep 10 & IPOL &  0.27 x 0.21 & $-$29 &  354 &  \n &
0.7 &   197&  $-$0.9, 0.9, 1.8, 3.6, 7.2, 14.4, 28.8, 57.6, 95\\
    &     &   PPOL &       \n     &  \n &  $<2.0$ &  20 & 0.5 &       4.0&\n  \\
1803+784  & 1999 Apr 6 &  IPOL & 0.20 x 0.17 & 39 & 1863 & \n & 1.4 &
 1165 & $-0.4$, 0.4, 0.8, 1.6, 3.2, 6.4, 12.8,  25.6 \\
    &     &   PPOL &     \n       & \n & 162 & 350 & 1.0 & 71 & \n \\
1807+698  & 1999 Apr 6 &  IPOL & 0.29 x 0.24 & 63 & 735 & \n & 0.8 &
 381 & $-0.6$, 0.6, 1.2, 2.4, 4.8, 9.6, 19.2, 38.4, 76.8 \\
    &     &   PPOL &     \n       & \n & $<4.0$ & 700 & 0.8 &  5.0 & \n \\ 
1823+568 & 1998 Jul 31 & IPOL & 0.23 x 0.17 & $-21$ & 1413 & \n & 0.9
 & 1057 & $-0.3$, 0.3, 0.6, 1.2, 2.4, 4.8  \\
 &       &  PPOL &  \n              &  \n          & 119   & 123 & 1.0 & 90 & \n \\
1828+487 & 2000 Sep 10 & IPOL &  0.33 x 0.23 &  $-$5 & 1230 &  \n &
0.7 &   808&  $-$0.25, 0.25, 0.5, 1, 2, 4\\
    &     &   PPOL &       \n     &  \n &  47  & 31  & 0.5 &     22&\n \\
1928+738\tablenotemark{a}&  1999 Jan 12 & IPOL &  0.27 x 0.25 & $ 54  $&  2423 &\n  &
0.8 &  1284& $-0.25$, 0.25, 0.5, 1, 2, 4, 8, 16, 32, 64\\
 &       &  PPOL &  \n              &  \n          &  76 &  156 &  0.8 &    37 & \n \\
1954+513  & 1999 Apr 6 &  IPOL & 0.27 x 0.23 & $-10$ &   1005 & \n &
0.8 &  780 & $-0.3$, 0.3, 0.6, 1.2, 2.4, 4.8, 9.6, 19.2 \\
    &     &   PPOL &     \n       & \n & 32 & 70  & 0.8 &  32 & \n \\
2021+614  & 1999 Apr 6 &  IPOL & 0.30 x 0.25 & $-34$ & 1051 & \n & 0.7
&  456 & $-0.5$, 0.5, 1, 2, 4, 8, 16, 32, 64 \\
    &     &   PPOL &     \n       & \n  &$ <3.0$ & 70 & 0.6 &  4.0 & \n \\
2200+420 & 1998 Jul 31 & IPOL & 0.31 x 0.19 & $-4$ & 5303 & \n & 1.1 &
 3718 & $-0.1$, 0.1, 0.2, 0.4, 0.8\\
  &       &  PPOL &  \n              &  \n          & 214 & 123& 0.8 &     75 &  \n \\
2351+456 & 2000 Sep 10 & IPOL &  0.41 x 0.19 &  20 & 1362 & \n  & 1.1
& 1008& $-$0.35, 0.35, 0.7, 1.4, 2.8  \\
    &     &   PPOL &       \n     &  \n &  24  & 51 & 0.6 &    18&\n \\
\enddata

\tablecomments{Columns are as follows: (1) Source
name. (2) UT Observation date. (3) Image polarization type.  (4) FWHM
dimensions of Gaussian restoring beam, in mas. (5) Position angle of
restoring beam, in degrees. (6) Total cleaned flux density [mJy]. (7)
Electric vector scaling in image [$\rm mJy \; mas^{-1}$].  (8) Dynamic range
(peak/RMS). (9) Peak intensity [$\rm mJy \; beam^{-1}$]. (10) Contour
levels, expressed as a percentage of peak intensity. }

\tablenotetext{a}{Image presented in \cite{LS00}}
\end{deluxetable}

\clearpage
\begin{deluxetable}{lrrrrcrrrr}
\tabletypesize{\scriptsize}
\tablecolumns{10}
\tablecaption{\label{coreprops}Properties of Pearson-Readhead VLBI
cores at 43 GHz}
\tablewidth{0pt}
\tablehead{ \colhead{Source} & \colhead{$S$} &   \colhead{$\log{L_{core}}$}
& \colhead{$m_{core}$}   &  \colhead{EVPA}  &
\colhead{$|{ EVPA} - { JPA}|$} & \colhead{$R$} & \colhead{Maj.} &\colhead{a}
&\colhead{PA} \\
\colhead{ [1]} & \colhead{ [2]} & \colhead{ [3]}  &
\colhead{ [4]} & \colhead{ [5]} & \colhead{ [6]} & \colhead{ [7]} 
&\colhead{ [8]} & \colhead{ [9]} & \colhead{ [10]} }
\startdata
\sidehead{FS-PR sample objects}
0016+731 &   616 &27.71& 2.0&  89&  43  & 5.23& 0.06&0.44&83 \\
0133+476 &  4016 &27.87& 4.9& $-$36&   6& 5.46&0.07&0.67&2 \\
0212+735 &   467 &27.84& 5.1& $-33$&  26& 0.23&0.09&0.18&94 \\
0316+413 &   169 &23.10&$<$3.0&  \n&  \n& 0.04&0.26&0.57&77 \\
0454+844 &    77 &24.39&$<$2.5&  \n&  \n& 2.44&0.00&\n&\n \\
0723+679 &   677 &27.08& 2.0&  52&  24&11.86&0.10&0.43&72 \\
0804+499 &   704 &27.57& 1.6& $-$88&  35&20.34&0.03&0.24&135 \\
0814+425 &   750 &26.05& 3.8& $-$80&   4& 5.62&0.07&0.44 & 43 \\
0836+710 &  1114 &28.14& 1.6& $-$80&  79& 1.38&0.07&0.27&56 \\
0850+581 &   190 &26.93&$<$2.2&  \n&  \n& 1.10& 0.00&\n&\n \\
0859+470 &   262 &27.16& 3.3&  30&  33& 1.00& 0.00&\n&\n \\
0906+430 &   363 &26.61&$<$0.7&  \n&  \n& 1.78&0.13&0.37&146 \\
0923+392 &   298 &26.56&$<$1.6&  \n&  \n& 0.03&0.06&0.23&108 \\
0945+408 &   465 &27.27& 5.5& $-$46&   3& 0.34& 0.00&\n&\n \\
0954+658 &   402 &26.13& 1.2& $-$58&  9& 6.81&0.02&0.74&108 \\
1624+416 &   206 &27.55&$<$2.2&  \n&  \n& 0.56& 0.00&\n&\n \\
1633+382 &  1754 &28.17& 2.5&  24&  75& 1.44&0.09&0.31&1 \\
1637+574 &   800 &27.05& 1.0& $-$54&  74& 0.47&0.04&0.23&21 \\
1641+399 &  6234 &27.74& 5.0&  81&  23& 2.22&0.14&0.16&89\\
1642+690 &   892 &27.10& 6.0& $-$14&   8& 1.68 & 0.00&\n&\n\\
1652+398 &   278 &23.88& 1.2&  27&  47& 1.48&0.13&0.86&95 \\
1739+522 &   645 &27.50& 1.9&  34&  10& 2.35&0.08&0.55&177 \\
1749+701 &   199 &26.47&$<$1.0&  \n&  \n& 0.81&0.05&0.56&119 \\
1803+784 &  1242 &27.16& 5.6& $-$45&  24& 1.32&0.06&0.82&162 \\
1807+698 &   326 &24.32&$<$1.1&  \n&  \n& 0.77&0.07&0.29&78 \\
1823+568 &  1065 &27.07& 8.5&  13&   8& 2.04&0.06&0.44&169 \\
1928+738 &   298 &25.83&$<$0.8&  \n&  \n& 0.11&0.03&0.73&69 \\
1954+513 &   774 &27.47& 4.0& $-$52&   1& 1.77&0.07&0.34&138 \\
2021+614 &    12 &24.20& $<$25.0&  \n&  \n& 0.01& 0.00&\n&\n \\
2200+420 &  4123 &25.69& 1.8&  73&  44& 3.31&0.07&0.22&40 \\
2351+456 &  1209 &28.10& 1.1&  77&  64& 3.29&0.13&0.46&95 \\
\sidehead{Other PR sample objects}
0153+744 &    84 &27.08&$<$2.5&  \n&  \n&10.68&0.07&0.16&35 \\
0538+498 &   177 &26.12&$<$3.2&  \n&  \n& 1.15&0.05&0.51&99 \\
0711+356 &    73 &26.69&$<$2.8&  \n&  \n& 0.69& 0.00&\n&\n \\
1828+487 &   649 &26.89& 2.1& $-$72&  23& 0.73&0.09&0.19&68 \\

\enddata 
\tablecomments{\scriptsize  All derived quantities are calculated assuming
$h = 0.65$, $q_o = 0.1$, and zero cosmological constant.\\
Col. (1).---  IAU source name. \\
Col. (2).---  Flux density of core component [mJy]. \\
Col. (3).---  Luminosity of core component  [$\rm W\; Hz^{-1}$]. \\
Col. (4).---  Fractional polarization of core component [$\%$].  \\
Col. (5).---  Electric vector position angle of core component [degrees]. \\
Col. (6).---  Offset of electric vector position angle of core with
respect to the innermost jet direction. \\
Col. (7).---  Ratio of VLBI core-to-remaining VLBI flux density
(source frame). \\
Col. (8).--- FWHM major axis of fitted Gaussian component in milliarcseconds. 
 A value of zero denotes a delta-function component. \\ 
Col. (9).--- Axial ratio of fitted component.  \\ 
Col. (10).--- Position angle of component's major axis.  \\ 
}
\end{deluxetable}

\clearpage
\begin{deluxetable}{lcrrrrrrrrrrr}
\tabletypesize{\scriptsize}
\tablecolumns{13}
\tablecaption{\label{jetprops}Jet Component Properties}
\tablewidth{0pt}
\tablehead{ \colhead{Source} & \colhead{Name} &  \colhead{$d$} & \colhead{$PA$} & \colhead{$S$}
& \colhead{$\log{L}$} & \colhead{$m$} & \colhead{EVPA} & \colhead{${JPA}_{in}$} & \colhead{${ JPA}_{out}$} 
& \colhead{Maj.} & \colhead{a} & \colhead{PA}  \\
\colhead{ [1]} & \colhead{ [2]} & \colhead{ [3]}  &
\colhead{ [4]} & \colhead{ [5]} & \colhead{ [6]} &
\colhead{ [7]}& \colhead{ [8]}  &  \colhead{ [9]}  & \colhead{ [10]} &
\colhead{ [11]}& \colhead{ [12]}  &  \colhead{ [13]} 
}
\startdata
\sidehead{FS-PR sample objects}
0016+731&A&    0.93&  132&     50&   26.97&$<$36&\n&  132&  105&0.74&0.45&115 \\
\\
0133+476&B&    0.09&  330&    384&   27.07&   5.4&   143&  330&  331&0.27&0.37&174 \\*
&A&    0.53&  331&     41&   26.10&$<$19&\n&  331&  331&0.30&0.50&52 \\*
\\
0212+735&D&    0.21&  121&    231&   27.96&$<$2.8&\n&  121&   98 &  0.16 & 0.64 &115 \\*
&C&    0.51&  107&     36&   27.15&  11.0&   111&   98&  138 & 0.00 & \n & \n  \\*
&B&    0.67&  115&    223&   27.94&$<$7.3&\n&  138&  159 &0.44 & 0.43 & 65 \\*
&A&    0.86&  126&    267&   28.02&  17.4&   133&  159&\n& 0.35 & 0.45 & 70 \\*
\\
0316+413&D&    0.55&  176&   1165&   23.94&$<$0.6&\n&  176&  188&0.20&0.32&155 \\*
&C&    0.95&  182&    783&   23.77&$<$1.3&\n&  188&  257&0.56&0.10&38 \\*
&B&    1.04&  195&    765&   23.76&$<$1.4&\n&  257&  231&0.45&0.36&63 \\*
&A&    1.31&  203&    734&   23.74&$<$1.0&\n&  231&  212&0.23&0.87&5 \\*
\\
0454+844&B&    0.31&  177&     14&   23.68&$<$13&\n&  177&  308 & 0.00 & \n & \n \\*
&A&    0.90&  144&      7&   23.38&\n&\n&  308&\n & 0.65 & 0.12 & 145 \\*
\\
0723+679&A&    0.75&  256&     30&   25.94&$<$19&\n&  256&\n&0.06&1.00 & 173 \\*
\\
0804+499&A&    0.32&  127&     20&   26.33&$<$17&\n&  127&\n&0.92&0.16&103 \\*
\\
0814+425&B&    0.35&  103&     34&   24.79&  12.9& 84&  103&  103 & 0.29 & 0.61 & 96 \\*
&A&    1.35&   86&     67&   25.08&$<$12&\n&   81&  167 & 0.40 & 0.67 & 65 \\*
\\
0836+710&C&    0.19&  201&    175&   27.74&   2.2&   171&  201&  225&0.11&0.14&50 \\*
&B&    0.36&  212&     75&   27.37&$<$3.8&\n&  225&  213&0.26&0.48&77 \\*
&A&    2.84&  213&     35&   27.04&\n&\n&  213&\n&0.58&0.83&31 \\*
\\
0850+581&A&    0.17&  227&     53&   26.67&  $<5.6$&     \n&  227&  153& 0.00 & \n & \n  \\*
\\
0859+470&A&    0.45&  357&    92&   27.02&  7.4&   157& 357&  336& 0.66 & 0.15 & 156  \\*
\\
0906+430&B&    0.32&  151&     95&   26.21&   7.5&   146&  151&  152&0.26&0.51&139 \\*
&A&    1.29&  153&     19&   25.51&$<$36&\n&  152&  152&0.97&0.17&153 \\*
\\
0923+392&e&    1.05&   93&     36&   25.83&      $<$40&\n&   93&  123& 0.28 & 0.50 & 150   \\*
&d&    1.38&  101&     48&   25.95&$<$21&\n&  123&  135 & 0.26 & 0.16 & 164 \\*
&C&    1.79&  109&    102&   26.28&$<$15&\n&  135&   77 & 0.33 & 0.45 & 152  \\*
&b&    2.17&  103&     69&   26.11&$<$9.7&\n&   77&   94 & 0.00 & \n & \n  \\*
&A&    2.83&  101&   2027&   27.58&   5.7&    71&   94&   81 & 0.00 & \n & \n  \\*
\\
0945+408&C&    0.13&  137&    455&   27.54&  5.9&  154& 137&  122& 0.20 & 0.18 & 134 \\*
&B&    0.44&  127&    113&   26.94&$<$14&\n&  122&  262& 0.35 & 0.36 & 142  \\*
&A&    0.66&  110&     85&   26.81&$<$12&\n&  262&  108& 0.34 & 0.49 & 176 \\*
\\
0954+658&B&    0.23&  311&      8&   24.54&$<$8.3&\n&  311&  337&0.00&\n&\n \\*
&A&    0.48&  326&     13&   24.75&$<$37&\n&  337&0&0.00&\n&\n \\*
\\
1624+416&D&    0.19&  262&     62&   27.47&   6.1&   144&  262&  262&0.18&0.48&99\\*
&C&    0.74&  262&     10&   26.68&\n&\n&  262&  237&0.00&\n&\n  \\*
&B&    1.42&  250&     49&   27.37&$<$38&\n&  237&  219&0.85&0.21&160 \\*
&A&    2.17&  239&      7&   26.52&\n&\n&  219&0.00 &0.00&\n&\n \\*
\\
1633+382&c2&    0.10&  279&     92&   27.25&   2.0&    21&  279&  264 & 0.28 & 0.43 &99 \\*
&c1&    0.47&  267&     41&   26.90&   8.7&   116&  264&  317 & 0.09 & 0.47 & 153  \\*
&B&    0.81&  291&     50&   26.99&  $<$16&\n&  360&  268 & 0.36 & 0.28 & 167 \\*
&A&    2.02&  269&     49&   26.98&  $<$18&\n&  268&  288 & 0.32 & 0.67 & 126 \\*
\\
1637+574&B&    0.23&  200&    988&   27.34&   3.8&   113&  200&  205&0.09&0.58&44 \\*
&A&    1.37&  204&     71&   26.20&   9.9&    24&  205&  200&0.20&0.83&64 \\*
\\
1641+399&G&    0.24&  284&    144&   26.27&   5.0&    89&  284&  262&0.00&\n&\n\\*
&F&    0.45&  274&     30&   25.59&  14.7&    63&  262&  210&0.00&\n&\n \\*
&E&    0.66&  247&    547&   26.85&   8.1&    39&  210&  210&0.34&0.67&32\\*
&D&    0.81&  240&    175&   26.35&  11.3&    32&  210&  294&0.00&\n&\n \\*
&C&    1.29&  256&    122&   26.20&   7.0&    53&  263&  298&0.00&\n&\n \\*
&B&    1.44&  261&    395&   26.71&   2.7&   150&  300&  300&0.33&0.62&145\\*
&A&    2.56&  260&    146&   26.27&  27.9&    37&  247&  247&0.46&0.64&114 \\*
\\
1642+690&C&    0.07&  158&    257&   26.76&   6.7&   169&  158&  171&0.14&0.70&176\\*
&B&    0.42&  171&     31&   25.84&  19.8&   132&  171&  180&0.28&0.68&5 \\*
&A&    1.19&  177&     28&   25.79&  12.5&   173&  180&  205&0.24&0.18&24 \\*
\\
1652+398&A&    0.55&  160&     67&   23.28 & $<$33&\n&  160&  160& 0.95 & 0.55 & 163 \\*
\\
1739+522&A&    0.12&  204&     79&   26.89&   3.5&    22&  204&  126 &0.00&\n&\n \\*
\\
1749+701&B&    0.21&  275&     16&   25.58&$<$5.1&\n&  275&  329&0.00&\n&\n  \\*
&A&    0.42&  307&     82&   26.29&$<$5.1&\n&  329&  277&0.29&0.39&158 \\*
\\
1803+784
&C&    0.22&  291&    307&   26.73&  16.0&   123&  291&  270&0.27&0.46&72 \\*
&B&    0.51&  279&    169&   26.47&  39.0&    94&  270&  284&0.72&0.33&120 \\*
&A&    1.46&  270&    140&   26.39&$<$18&\n&  236&  236&0.63&0.47&58 \\*
\\
1807+698&A&    0.21&  252&    190&   24.10&$<$1.5&\n&  252&  257&0.14&0.61&83 \\*
\\
1823+568&C&    0.12&  201&    109&   26.26&  13.3&    21&  201&  195&0.00&\n&\n \\*
&B&    0.48&  196&    148&   26.39&  21.6&    11&  195&  201&0.13&0.56&173 \\*
&A&    1.52&  200&     48&   25.90&  41.7&    41&  201&  226&0.32&0.44&56 \\*
\\
1928+738&G&    0.24&  166&   1338&   26.58&   2.7&   100&  166&  159 & 0.09 & 0.44 & 3 \\*
&f2&    0.54&  162&    209&   25.77&  $<$2.0&\n&  159&  168 & 0.14 & 0.73 & 150  \\*
&f1&    0.76&  164&     14&   24.60&   $<$8.5&\n&  168&  127 & 0.00 & \n & \n  \\*
&e2&    1.06&  153&     87&   25.39&   7.0&    82&  127&  159 & 0.18 & 0.55 & 147  \\*
&e1&    1.42&  155&    260&   25.87&   $<$3.4&\n&  159&  205 & 0.38 & 0.59 & 167  \\*
&D&    1.66&  164&     18&   24.71&  $<$14&\n&  205&  149 & 0.00 & \n &\n \\*
&C&    2.77&  158&     38&   25.03&  $<$18&\n&  149&  254 & 0.35 & 0.41 & 55  \\*
&B&    2.74&  168&      8&   24.35&  28&    71&  254&  171 &0.00 & \n &\n \\*
&A&    3.20&  168&     11&   24.49&  19&    41&  171&  311 & 0.00& \n & \n \\*
\\
1954+513&B&    0.20&  307&    135&   26.99&   3.6&   120&  307&  307&0.29&0.33&129 \\*
&A&    0.91&  306&     59&   26.63&$<$14&\n&  307&  307&0.36&0.38&135 \\*
\\
2021+614&B&    4.04&   33&     45&   24.84&$<$20&\n&   33&\n&0.57&0.49&148 \\*
&C&    2.30&  214&    115&   25.25&$<$5&\n&  214&  201&0.42&0.08&177 \\*
&D1&    3.03&  211&    594&   25.96&$<$0.7&\n&  201&  222 &0.35&0.38&15\\*
&D2&    3.15&  212&    262&   25.61&$<$0.7&\n&  222&\n&0.29&0.16&125 \\*
\\
2200+420&C&    0.19&  209&    686&   24.94&   3.3&    46&  209&  208&0.32&0.17&20 \\*
&B&    1.98&  193&    186&   24.37&  25.3&    23&  208&  191&0.29&0.83&73 \\*
&A&    2.38&  195&    298&   24.58&  21.6&    37&  191&  204&0.38&0.66&116 \\*
\\
2351+456&A&    0.22&  321&     98&   27.38&   4.5&   144&  321&  296&0.32&0.26&36\\*
\\
\sidehead{Other PR sample objects}
0153+744&A&    0.49&   68&      3&   26.06&\n&\n&   68&0.00 &0.00 & \n &\n  \\
\\
0711+356&B&    0.50&  329&      7&   26.01&$<$29&\n&  329&  337& 0.00 & \n &\n  \\*
&A&    6.30&  337&     41&   26.78&$<$18&\n&  337&\n&0.53&0.86&40 \\*
\\
1828+487&E&    0.09&  311&    322&   26.77&   7.4&   106&  311&  319&0.23&0.51&113 \\*
&D&    0.81&  319&    124&   26.36&   5.7&   133&  319&  341&0.52&0.30&161 \\*
&C&    1.85&  333&     22&   25.61&$<$31&\n&  341&  331&0.71&0.15&142 \\*
&B&    4.25&  332&     32&   25.77&  24.4&   163&  331&  325&0.53&0.31&149 \\*
&A&   10.23&  328&     65&   26.08&  \n&   \n&  325&  272&1.56&0.47&120 \\*
\\

\enddata 

\tablecomments{\footnotesize  All luminosities are calculated assuming $h = 0.65$, $q_o = 0.1$, and zero cosmological constant. \\
Col. (1).--- IAU source name. \\
Col. (2).--- Component name. \\
Col. (3).--- Distance from core component in milliarcseconds. \\
Col. (4).--- Position angle with respect to the core
component. \\
Col. (5).---  Fitted Gaussian flux density in mJy. \\
Col. (6).---  Luminosity in $\rm W \; Hz^{-1}$. \\
Col. (7).---  Fractional linear polarization in per cent, measured at
position of I peak. \\
Col. (8).---  Electric vector position angle, measured at
position of I peak. \\
Col. (9).---  Local jet direction, measured upstream of the
component. \\
Col. (10).--- Local jet direction, measured downstream of the
component. \\ 
Col. (11).--- FWHM major axis of fitted component in milliarcseconds. 
 A value of zero denotes a delta-function component. \\ 
Col. (12).--- Axial ratio of fitted component.  \\ 
Col. (13).--- Position angle of component's major axis.  \\ 
}
\end{deluxetable}

\clearpage
\begin{deluxetable}{lll}
\tabletypesize{\scriptsize}
\tablecolumns{3}
\tablecaption{\label{quantities}Measured and Derived Quantities for the Pearson-Readhead AGN Sample}
\tablewidth{0pt}
\tablehead{\colhead{Symbol} & \colhead{Property} &\colhead{Units} }
\startdata
\sidehead{General Properties}
\n &	Optical classification (RG, LPRQ, HPQ, BL Lac) & \n \\
$z$ & Redshift & \n \\
$\bar \alpha_{5-15}$  &   Time-averaged spectral index between 4.8 and 14.5 GHz &\n  \\

\sidehead{Variability Properties}
$V_5$ & Variability amplitude at 5 GHz &\n \\
$V_{15}$ & Variability amplitude at 14.5 GHz &\n \\
$T_{b,var}$ & Variability brightness temperature & K   \\
 
\sidehead{Luminosity Properties} 
$L_{T,5}$ &  Total single-dish luminosity at 5 GHz &$\rm W\; Hz^{-1}$  \\
$L_{t,43}$ &  Total VLBI luminosity at 43 GHz &$\rm W\; Hz^{-1}$  \\
$L_{c,5}$ & VLBI core luminosity at 5 GHz &  $\rm W\; Hz^{-1}$  \\
$L_{c,43}$ & VLBI core luminosity at 43 GHz &  $\rm W\; Hz^{-1}$  \\
$L_{C,1.4} $ & Kiloparsec-scale (VLA) core luminosity at 1.4 GHz &  \\
$L_{opt}$  & Optical (V-band) luminosity &  $\rm W\; Hz^{-1}$ \\
$L_{xray}$ &  X-ray luminosity at 1 keV & $\rm W\; Hz^{-1}$  \\

\sidehead{Jet Properties} 
$R$ & Ratio of core to remaining VLBI flux density at 43 GHz & \n  \\
$F_c$ & Ratio of VLBI core to remaining total flux density at 5 GHz & \n \\
$N_{bend}$ &  Number of significant bends ($>10\arcdeg$) within  $100\; h^{-1}$ projected pc of the core.& Deg. \\
$\Sigma_{bend}$ & Sum of significant bend angles within  $100\; h^{-1}$ projected pc of the core& Deg. \\
$\beta_{app}$ & Fastest measured component speed in jet in units of $c$ & \n \\ 
$L_{cpt}$ & Luminosity of jet component at 43 GHz  &$\rm W\; Hz^{-1}$  \\
$d_{cpt}$ & Projected distance of jet component from core & Pc  \\

\sidehead{Polarization Properties}
$m_{cpt}$ &   Linear polarization of jet component at 43 GHz &  $\%$ \\
$m_{c,43}$  &Linear polarization of VLBI core at 43 GHz &  $\%$ \\
$m_{t,43}$&   Total VLBI linear polarization of source at 43 GHz  &  $\%$ \\
$m_{opt}$ &	Optical V band linear polarization & $\%$ \\
${ EVPA}$ & Electric vector position angle at 43 GHz &   Deg. \\
$|{ EVPA}_{core} - { JPA}|$  & Offset of core ${ EVPA}$ from innermost jet direction & Deg. \\
$|{ EVPA}_{cpt} - { JPA}_{in}|$  & Offset of component ${ EVPA}$ from upstream jet direction & Deg. \\
$|{ EVPA}_{cpt} - { JPA}_{out}|$  & Offset of component ${ EVPA}$ from downstream jet direction & Deg. \\

\enddata 
 \end{deluxetable}

\clearpage
\begin{deluxetable}{llrrlrl}
\tabletypesize{\scriptsize}
\tablecolumns{7}
\tablecaption{\label{corr_results}Correlations Involving VLBI Properties of the FS-PR Sample}
\tablewidth{0pt}
\tablehead{\colhead{Property} & \colhead{Property} & \colhead{N} & \colhead{$\tau$} & \colhead{P}
& \colhead{$\tau_z$} & \colhead{$P_z$}  \\
\colhead{(1)} & \colhead{(2)} &\colhead{(3)} & \colhead{(4)} &\colhead{(5)} & \colhead{(6)} &\colhead{(7)}}
\startdata
$m_{t,43}$ &$\bar \alpha_{5-15}$   &   30 & 0.425 &$8.79 \times 10^{-2}$ & 0.429   &$ 2.97 \times 10^{-4}$ \\
$R$   & $m_{opt}$   &   29 & 0.429 &$1.09 \times 10^{-1}$ & 0.444        &$ 3.15 \times 10^{-3}$ \\
$m_{c,43} $& $m_{t,43}$& 31 & 0.445 &$4.58\times 10^{-3}$ & 0.447                   &$ 4.19 \times 10^{-3}$\\
$m_{c,43}$ & $\bar \alpha_{5-15}$ & 30 & 0.338 &$3.26 \times 10^{-1}$ & 0.346   &$ 4.64 \times 10^{-3}$ \\
$m_{c,43}$ & $|{ EVPA}_{core} - {JPA}|$ & 21 &$-0.452$ &$4.04 \times 10^{-1}$ &$-0.459$&$ 7.65 \times 10^{-3}$ \\
$R$         &$V_{5}$     &          30 & 0.358 &$5.47 \times 10^{-1}$ & 0.367      &$8.27 \times 10^{-2}$ \\
$m_{t,43}$  & $V_{15}$        &        30 & 0.329 &1.02 &  0.330              &$ 1.50 \times 10^{-1}$ \\
$N_{bend}$ & $\beta_{app}$ &         20 & 0.445 &$6.13 \times 10^{-1}$ & 0.371      &$ 6.21 \times 10^{-1}$ \\
$R$&        $T_{b,var}$ &     19 & 0.427 &1.07  & 0.457                             &$ 7.22 \times 10^{-1}$ \\
$R$ &$F_c$ &                                24 &  0.388&$7.84 \times 10^{-1}$ & 0.388& $7.63\times 10^{-1}$\\
$N_{bend}$ &   $L_{xray}$ &   26 &  0.359&1.01 &    0.204& 1.31 \\
$\Sigma_{bend}$ &$\beta_{app}$ &   20  & 0.353& $2.96$ &   0.312 & $2.79$ \\

\sidehead{Luminosity-luminosity correlations}
$L_{c,43}$&     $  L_{c,5} $& 24  &0.795 &$5.27\times 10^{-6}$ &  0.584 &$5.55\times 10^{-2}$\\
$L_{t,43}$ &$L_{C,1.4}$ &    29  &0.783 &$2.45\times 10^{-7}$ &  0.562 &$9.87 \times 10^{-2}$\\
$L_{t,43}  $&   $L_{xray}$ &    25 & 0.773 &$6.02\times 10^{-6}$ &  0.553 &$1.60 \times 10^{-1}$\\
$L_{t,43}$ &       $ L_{T,5}   $&  30  &0.798 &$5.98\times 10^{-8}$ &  0.585 &$1.69\times 10^{-1}$\\
$L_{c,43} $& $ L_{t,43} $&31& 0.784 &$5.89\times 10^{-8}$ &  0.589 &$2.16\times 10^{-1}$\\
$L_{c,43} $&     $L_{xray}$ &   25 & 0.725 &$3.84\times 10^{-5}$ &  0.413 &$6.08 \times 10^{-1}$\\
$L_{t,43} $& $L_{c,5}$&        24  &0.674 &$3.96\times 10^{-4}$ &  0.324 &1.25\\
$L_{c,43}$ &  $L_{opt}$&     31 & 0.590 &$3.16\times 10^{-4}$ &  0.326 &1.29\\

\sidehead{Correlations with redshift}
$L_{T,5}$  &  $z$ &         30 & 0.779 &  $1.47 \times 10^{-7}$ \\
$L_{C,1.4}$   & $z$ &       30 & 0.773 &  $1.96 \times 10^{-7}$ \\
$L_{c,43}$     &  $z$ &     31 & 0.706 &  $2.39 \times 10^{-6}$ \\
$L_{t,43}$      &  $z$ &    31 & 0.673 &  $1.04 \times 10^{-5}$ \\
$L_{xray}$      & $z$ &     26 & 0.717 &  $2.81 \times 10^{-5}$ \\
$L_{c,5}$    &  $z$ &       24 & 0.746 &  $3.23 \times 10^{-5}$ \\
$L_{opt}$  &   $z$ &        32 & 0.568 &  $4.86 \times 10^{-4}$ \\
$T_{b,var}$    &  $z$ &     20 & 0.589 &  $2.79 \times 10^{-2}$ \\

\sidehead{Correlations likely due to redshift bias}
$L_{t,43}$   &       $L_{opt}$    &     31 & 0.539& $2.02 \times 10^{-3}$&  0.260 & 2.15 \\
$L_{c,43}$   &       $L_{T,5}$    &       30 & 0.723& $2.04 \times 10^{-6}$&  0.391 & 3.19 \\
$L_{opt}$    &       $\beta_{app}$ &  20 & 0.379& $1.95 \times 10^{-1}$&  0.268 & 4.40 \\
$L_{c,43}$   &       $T_{b,var}$   &  19 & 0.591& $4.10 \times 10^{-2}$&  0.237 & 5.20 \\
$L_{c,43}$   &       $L_{C,1.4}$  &     29 & 0.718& $4.62 \times 10^{-6}$&  0.370 & 5.68 \\
$L_{opt}$    &       $N_{bend}$    &  32 & 0.305&  1.42               &   0.199 & 7.83 \\
$L_{C,1.4}$  &       $N_{bend}$    &  30 & 0.336& $9.02 \times 10^{-1}$&  0.187 & 8.08 \\

\enddata 
\tablecomments{\scriptsize Columns are as follows: (1) and (2) Source property
(see Table~\ref{quantities} for symbol definitions); (3) Number of data pairs; (4) Kendall's
tau coefficient; (5) Probability of correlation arising by chance in
per cent; (6) Kendall's tau coefficient, with redshift partialed out;
(7) Probability of correlation arising by chance, with redshift
partialed out.}
 
 \end{deluxetable}

\clearpage
\begin{deluxetable}{llrr}
\tabletypesize{\scriptsize}
\tablecolumns{4}
\tablecaption{\label{ks_results}Two-Sample tests for various subclasses of the FS-PR sample}
\tablewidth{0pt}
\tablehead{& & \colhead{QSO} & \colhead{HPQ}    \\
\colhead{Source} & &  \colhead{vs.} &  \colhead{vs.}  \\
\colhead{Property} & \colhead{Symbol}&\colhead{BL Lacs}&\colhead{BL Lacs}  \\
\colhead{(1)} & \colhead{(2)}&\colhead{(3)}&\colhead{(4)}}
\startdata
\sidehead{\it General properties}
Redshift & $z$                               &{\bf 0.07}&  {\bf 1.87} \\ 
Total VLBI luminosity  &$L_{t,43}$   &{\bf 0.07}& {\bf 0.60}  \\ 
Total fractional VLBI polarization & $m_{t,43}$  & 48.65 & 35.18 \\ 

\sidehead{\it VLBI Core properties}
Luminosity  & $L_{c,43}$ & {\bf 0.11} &  {\bf 0.48} \\  
Core dominance & $R$   &     54.96 & 98.21  \\
Fractional polarization & $m_{c,43}$  &  74.55 & 29.91 \\
EVPA offset from innermost jet direction& $|{ EVPA}_{core} - { JPA}|$ & 86.66 & 99.38 \\

\sidehead{\it Jet properties}
Component luminosity & $L_{cpt}$  &  {\bf 0.07}&  {\bf 0.06}\\
Component fractional polarization & $m_{cpt}$  &  2.96 & 21.86\\
Component EVPA offset from upstream jet direction & $|{ EVPA}_{cpt} - { JPA}_{in}|$  & 16.47 & 37.30 \\
Component EVPA offset from downstream jet direction & $|{ EVPA}_{cpt} - { JPA}_{out}|$ & {\bf 1.12}& 2.52 \\
Component Distance from core & $d_{cpt}$  &    {\bf   0.20} &    2.97  \\
Number of jet bends & $N_{bend}$ &53.38 & 60.30 \\
Sum of bend angles & $\Sigma_{bend}$ &   87.19 &     60.30 \\  
Maximum apparent speed & $\beta_{app}$ &   71.35 &    44.279 \\

\enddata 

\tablecomments{All values represent the probability (in per cent) that the subsamples
were drawn from the same population, based on the source property in
question. Values less than $2\%$ (in bold) are considered
statistically significant. Columns are as follows: (1) Source
property; (2) Symbol; (3) All quasars vs. BL Lacs; (4) High-optical
polarization quasars vs. BL Lacs.}
\end{deluxetable}

\end{document}